\documentclass[prd,nofootinbib,showpacs,preprint]{revtex4-1}
\usepackage[T1]{fontenc}
\usepackage[usenames,dvipsnames]{color}
\usepackage[colorlinks,citecolor=blue]{hyperref}
\usepackage{amsmath,amssymb,epsfig,url,graphicx,slashed,cancel,epstopdf,multirow,graphics,setspace,tikz-feynhand}
\newcommand{\be}{\begin{equation}}
\newcommand{\ee}{\end{equation}}
\newcommand{\bea}{\begin{eqnarray}}
\newcommand{\eea}{\end{eqnarray}}
\newcommand{\mathsym}[1]{{}}
\newcommand{\unicode}[1]{{}}
\begin{document}
\preprint{}
\title{Neutrinos, vacuum stability and triple Higgs coupling in SMASH}
\author{Timo J. K\"{a}rkk\"{a}inen}
\email{karkkainen@caesar.elte.hu}
\affiliation{Institute for Theoretical Physics, ELTE Eötvös Loránd University, Pázmány Péter sétány 1/A, 1117 Budapest, Hungary}
\author{Katri Huitu}
\email{katri.huitu@helsinki.fi}
\affiliation{Department of Physics and Helsinki Institute of Physics, P. O. Box 64, FI-00014 University of Helsinki, Finland}
\author{C.R. Das}
\email{das@theor.jinr.ru}
\affiliation{Bogoliubov Laboratory of Theoretical Physics, Joint Institute for Nuclear Research, International Intergovernmental Organization, Joliot-Curie 6, 141980 Dubna, Moscow region, Russian Federation}
\date{\today}

%%%%%%%%%%%%%%%%%%%%%%

\begin{abstract}

We perform a phenomenological analysis of the observable consequences on the extended scalar sector of the SMASH (Standard Model - Axion - Seesaw - Higgs portal inflation) framework. We solve the vacuum metastability problem in a suitable region of SMASH scalar parameter spaces and discuss the one-loop correction to triple Higgs coupling $\lambda_{HHH}$. We also find that the correct neutrino masses and mass squared differences and baryonic asymmetry of the universe can arise from this model and consider running of the Yukawa couplings of the model. In fact, we perform a full two-loop renormalization group analysis of the SMASH model.

\end{abstract}

%%%%%%%%%%%%%%%%%%%%%%

%$a_0$
\maketitle

%%%%%%%%%%%%%%%%%%%%%%

\section{Introduction}

After the discovery of the Standard Model (SM) Higgs boson \cite{Aad:2012tfa,Chatrchyan:2012xdj}, every elementary particle of the SM has been confirmed to exist. Even though the past forty years have been a spectacular triumph for the SM, the mass of the Higgs boson ($m_H = 125.09 \pm 0.32$ GeV) poses a serious problem for the SM. It is well-known that the SM Higgs potential is metastable \cite{Alekhin2012}, as the sign of the quartic coupling, $\lambda_H$, turns negative at instability scale $\Lambda_\text{IS} \sim 10^{11}$ GeV. On the other hand, the SM is devoid of nonperturbativity problem, since the nonperturbativity scale $\Lambda_\text{NS} \gg M_{Pl}$, where $M_{Pl} = 1.22\times 10^{19}$ GeV is the Planck scale. At the post-Planckian regime effects of quantum gravity are expected to dominate, and the nonperturbativity scale is therefore well beyond the validity region of the SM, unlike the instability scale. The largest uncertainties of SM vacuum stability are driven by top quark pole mass and the mass of SM Higgs boson. The current data is in significant tension with the stability hypothesis, making it more likely that the universe is in a false vacuum state. The expected lifetime of vacuum decay to a true vacuum is extraordinarily long, and it is unlikely to affect the evolution of the universe. However, it is unclear why the vacuum state entered into a false vacuum, to begin with during the early universe. In this post-SM era, the emergence of vacuum stability problem (among many others) forces the particle theorists to expand the SM in such a way that the $\lambda_H$ will stay positive during the running all the way up to the Planck scale.

It is possible that at or below the instability scale heavy degrees of freedom originating from a theory beyond the SM start to alter the running of the SM parameters of renormalization group equations (RGE). This approach aims to solve the vacuum stability problem by proving that the universe is currently in a true vacuum. One theory candidate is a complex singlet scalar extended SM. The scalar sector of such a theory may stabilise the theory with a threshold mechanism \cite{Elias-Miro2012, Lebedev2012}. The effective SM Higgs coupling gains a positive correction $\delta \equiv \lambda_{H\sigma}^2/\lambda_{\sigma}$ at $m_\sigma$, where $\lambda_{H\sigma}$ is the Higgs doublet-singlet portal coupling an $\lambda_{\sigma}$ is the quartic coupling of the new scalar.

Corrections altering $\lambda_H$ would in such a model induce also corrections to triple Higgs coupling, $\lambda_{HHH}^\text{tree} = 3m_H/v^2$, where $v = 246.22$ GeV is the SM Higgs vacuum expectation value (VEV). The triple Higgs coupling is uniquely determined by the SM but unmeasured. In fact, the Run 2 data from Large Hadron Collider (LHC) have only been able to determine the upper limit of the coupling to be 15 times the SM prediction \cite{PDG18}. Therefore, future prospects of measuring a deviation of triple Higgs coupling by the high-luminosity upgrade of the LHC (HL-LHC) \cite{Cepeda:2019klc} or by a planned next-generation Future Circular Collider (FCC) \cite{Arkani-Hamed:2015vfh,Baglio:2015wcg,Contino:2016spe} gives us hints of the structure of the scalar sector of a beyond-the-SM theory. Previous work has shown that large corrections to triple Higgs coupling might originate from a theory with one extra Dirac neutrino \cite{Baglio:2016ijw}, inverse seesaw model \cite{Baglio:2016bop}, two Higgs doublet model \cite{Arhrib:2008jp,Dubinin:1998nt,Dubinin:2002nx}, one extra scalar singlet \cite{Kanemura:2015fra,Kanemura:2016lkz,He:2016sqr} or in Type II seesaw model \cite{Aoki:2012jj}.

The complex singlet scalar, and consequently the corresponding threshold mechanism, is embedded in a recent SMASH \cite{Smash1, Smash2, Smash3} theory, which utilizes it at $\lambda_{H\sigma} \sim -10^{-6}$ and $\lambda_{\sigma} \sim 10^{-10}$. The mechanism turns out to be dominant unless the new Yukawa couplings of SMASH are $\mathcal{O}(1)$. In addition to its simple scalar sector extension, SMASH includes electroweak singlet quarks $Q$ and $\overline{Q}$ and three heavy right-handed Majorana neutrinos $N_1$, $N_2$ and $N_3$ to generate masses to neutrinos.

The structure of this paper is as follows. In Sec. \ref{theory}, we summarize the SMASH model and cover the relevant details on its scalar and neutrino sectors. In Sec. \ref{methods}, we discuss the methods, numerical details, RGE running and our choice of benchmark points. Our results are presented of Sec. \ref{results}, where the viable parameter space is constrained by various current experimental limits. We give our short conclusions on Sec. \ref{conclusions}.

%%%%%%%%%%%%%%%%%%%%%%

\section{\label{theory}Theory}

\noindent SMASH framework \cite{Smash1, Smash2, Smash3} expands the scalar sector of the SM by introducing a complex singlet field 
\be \sigma = \frac{1}{\sqrt{2}}\left( v_\sigma + \rho \right) e^{iA/v_\sigma},
\ee
where $\rho$ and $A$ (the axion) are real scalar fields, and $v_\sigma \gg v$ is the VEV of the complex singlet. The scalar potential of SMASH is then
\be \label{eq:potential}
V(H,\sigma) = \lambda_H\left(H^\dagger H - \frac{v^2}{2}\right)^2 + \lambda_\sigma\left(|\sigma|^2- \frac{v_\sigma^2}{2}\right)^2 + 2\lambda_{H\sigma}\left(H^\dagger H - \frac{v^2}{2}\right)\left(|\sigma|^2- \frac{v_\sigma^2}{2}\right).
\ee
Defining $\phi_1 = H$ and $\phi_2 = \sigma$, the scalar mass matrix of this potential is
\bea
(M_{ij}) &= \frac{1}{2}\frac{\partial^2 V}{\partial \phi_i\partial \phi_j}\left|_{\substack{H = v/\sqrt{2},\\ \sigma = v_\sigma/\sqrt{2}}}\right. = 
\left( \begin{array}{cc}
	2\lambda_Hv^2 		& 2\lambda_{H\sigma}vv_\sigma\\
	2\lambda_{H\sigma}vv_\sigma & 	2\lambda_{\sigma} v_\sigma^2 
\end{array}\right),
\eea
which has eigenvalues
\bea
m_H^2 &\approx 2\left(\lambda_Hv^2 + \lambda_{H\sigma}v_\sigma^2\right),\\
m_\sigma^2 &\approx 2\left(\lambda_{\sigma}v_\sigma^2 +\lambda_{H\sigma}v^2\right).
\eea 
The SMASH framework also includes a new quark-like field $Q$, which has colour but is an electroweak singlet. It gains its mass via Higgs mechanism, through complex singlet $\sigma$. It arises from Yukawa term:
\be 
\mathcal{L}^Y_Q = Y_Q\overline{Q}\sigma Q \Rightarrow m_Q \approx \frac{Y_Qv_\sigma}{\sqrt{2}}.
\ee 
We will show later that $Y_Q = \mathcal{O}(1)$ is forbidden by vacuum stability requirement. The hypercharge of $Q$ is chosen to be $q = -1/3$, even though also $q=2/3$ is possible. Our analysis is almost independent of the hypercharge assignment.

%%%%%%%%%%%%%%%%%%%%%%

\subsection{Threshold correction}

Consider an energy scale below $m_\sigma < \Lambda_\text{IS}$, where the heavy scalar $\sigma$ is integrated out. The low-energy Higgs potential should match the SM Higgs potential:

\be
V(H) = \lambda_H^\text{SM}\left( H^\dagger H - \frac{v^2}{2}\right)^2.
\ee 
It turns out that the quartic coupling we measure has an additional term:
\be 
\lambda_H^\text{SM} = \lambda_H - \frac{\lambda_{H\sigma}^2}{\lambda_\sigma}.
\ee 
Since the SM Higgs quartic coupling will be approximately $\lambda_H(M_{Pl}) \approx -0.02$, the threshold correction
\be 
\delta \equiv \frac{\lambda_{H\sigma}^2}{\lambda_\sigma}
\ee 
should have a minimum value close to $|\lambda_H(M_{Pl})|$ or slightly larger to push the high-energy counterpart $\lambda_H$ to positive value all the way up to $M_{Pl}$. A too large correction will however increase $\lambda_H$ too rapidly, exceeding the perturbativity limit $\sqrt{4\pi}$. Similar to $\lambda_H$, the SM Higgs quadratic parameter $\mu_H$ gains a threshold correction:
\be 
\left(\mu_H^\text{SM}\right)^2 = \mu_H^2 -\frac{\lambda_{H\sigma}}{\lambda_{\sigma}}\mu_\sigma^2.
\ee 

In the literature \cite{Elias-Miro2012, Lebedev2012}, there are two possible ways of implementing this threshold mechanism. One may start by solving the SM RGE's up to $m_\sigma$, where the new singlet effects kick in, and the quadratic and quartic couplings gain sudden increments. Continuation of RGE analysis to even higher scales then requires utilizing the new RGE's up to the Planck scale.

Another way is to solve the new RGE's from the SM scale, not bothering to solve the low-energy SM RGE's at all. We will use the former approach.

\subsection{One-loop correction to triple Higgs coupling}

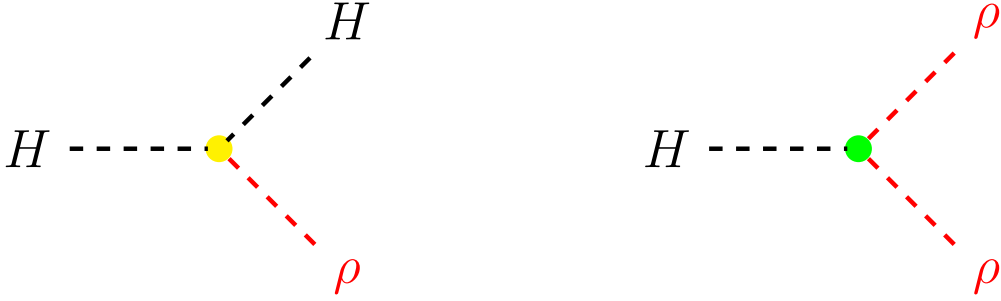
\begin{figure}\vspace{1cm}
\setlength{\feynhanddotsize}{1.75mm}
\setlength{\feynhandlinesize}{1pt}
\centering
\begin{align*}\hspace{-1.7cm}
\begin{tikzpicture}[transform canvas={scale=1.7}][baseline=(o.base)]
\begin{feynhand}
\vertex (a) at (-1.5,0) {$H$};
\vertex (b) at (1,-1) {\color{red}$\rho$};
\vertex (c) at (1,1) {$H$};
\vertex [dot, yellow] (o) at (0,0) {};
\propag [scalar, black] (a) to (o) {};
\propag [scalar, red] (b) to (o) {};
\propag [scalar, black] (c) to (o) {};
\hspace{1cm}$=- i\lambda_{H\sigma}v_{\sigma},$
\end{feynhand}
\end{tikzpicture}
\hspace{8.5cm}
\begin{tikzpicture}[transform canvas={scale=1.7}][baseline=(o.base)]
\begin{feynhand}
\vertex (a) at (-1.5,0) {$H$};
\vertex (b) at (1,-1) {\color{red}$\rho$};
\vertex (c) at (1,1) {\color{red}$\rho$};
\vertex [dot, green] (o) at (0,0) {};
\propag [scalar, black] (a) to (o);
\propag [scalar, red] (b) to (o);
\propag [scalar, red] (c) to (o);
\hspace{1cm}$=- i\lambda_{H\sigma}v$
\end{feynhand}
\end{tikzpicture}
\end{align*}
\vspace{1cm}
\caption{\label{fig:feynman} Vertex factors on trilinear vertices involving both SM Higgs boson and a real singlet $\rho$. They can be derived from Eq. (\ref{eq:potential}). We denote $\rho$ and its propagator by red colour.}
\end{figure}

\begin{figure}\vspace{2cm}
\setlength{\feynhandblobsize}{4mm}
\setlength{\feynhandlinesize}{2pt}
\centering
\begin{align*}
\begin{tikzpicture}[transform canvas={scale=0.7}][baseline=(o.base)]
\begin{feynhand}
\vertex (a) at (0,3.95) {\text{\Large{$H$}}};
\vertex [blob, yellow] (o1) at (0,1.8) {};
\propag [scalar, black] (a) to (o1) {};
\vertex [blob, black] (o2) at (-1.5,-0.75) {};
\propag [scalar, black] (o2) to [edge label = \text{\Large{$H$}}] (o1) {};
\vertex [blob, yellow] (o3) at (1.5,-0.75) {};
\propag [scalar, red] (o1) to [edge label = {\color{red}\text{\Large{$\rho$}}}] (o3) {};
\propag [scalar, black] (o3) to [edge label = \text{\Large{$H$}}] (o2) {};
\vertex (b) at (-3.5,-1.75) {\text{\Large{$H$}}};
\propag [scalar, black] (b) to (o2) {};
\vertex (c) at (3.5,-1.75) {\text{\Large{$H$}}};
\propag [scalar, black] (c) to (o3) {};
\end{feynhand}
\end{tikzpicture}
\hspace{5.5cm}
\begin{tikzpicture}[transform canvas={scale=0.7}][baseline=(o.base)]
\begin{feynhand}
\vertex (a) at (0,3.95) {\text{\Large{$H$}}};
\vertex [blob, yellow] (o1) at (0,1.8) {};
\propag [scalar, black] (a) to (o1) {};
\vertex [blob, yellow] (o2) at (-1.5,-0.75) {};
\propag [scalar, black] (o2) to [edge label = \text{\Large{$H$}}] (o1) {};
\vertex [blob, green] (o3) at (1.5,-0.75) {};
\propag [scalar, red] (o1) to [edge label = {\color{red}\text{\Large{$\rho$}}}] (o3) {};
\propag [scalar, red] (o3) to [edge label = {\color{red}\text{\Large{$\rho$}}}] (o2) {};
\vertex (b) at (-3.5,-1.75) {\text{\Large{$H$}}};
\propag [scalar, black] (b) to (o2) {};
\vertex (c) at (3.5,-1.75) {\text{\Large{$H$}}};
\propag [scalar, black] (c) to (o3) {};
\end{feynhand}
\end{tikzpicture}
\hspace{5.5cm}
\begin{tikzpicture}[transform canvas={scale=0.7}][baseline=(o.base)]
\begin{feynhand}
\vertex (a) at (0,3.95) {\text{\Large{$H$}}};
\vertex [blob, green] (o1) at (0,1.8) {};
\propag [scalar, black] (a) to (o1) {};
\vertex [blob, green] (o2) at (-1.5,-0.75) {};
\propag [scalar, red] (o2) to [edge label = {\color{red}\text{\Large{$\rho$}}}] (o1) {};
\vertex [blob, green] (o3) at (1.5,-0.75) {};
\propag [scalar, red] (o1) to [edge label = {\color{red}\text{\Large{$\rho$}}}] (o3) {};
\propag [scalar, red] (o3) to [edge label = {\color{red}\text{\Large{$\rho$}}}] (o2) {};
\vertex (b) at (-3.5,-1.75) {\text{\Large{$H$}}};
\propag [scalar, black] (b) to (o2) {};
\vertex (c) at (3.5,-1.75) {\text{\Large{$H$}}};
\propag [scalar, black] (c) to (o3) {};
\end{feynhand}
\end{tikzpicture}
\end{align*}
\vspace{0.5cm}
\caption{\label{fig:loop} One-loop corrections to SM triple Higgs coupling induced by the existence of an extra scalar singlet.}
\end{figure}
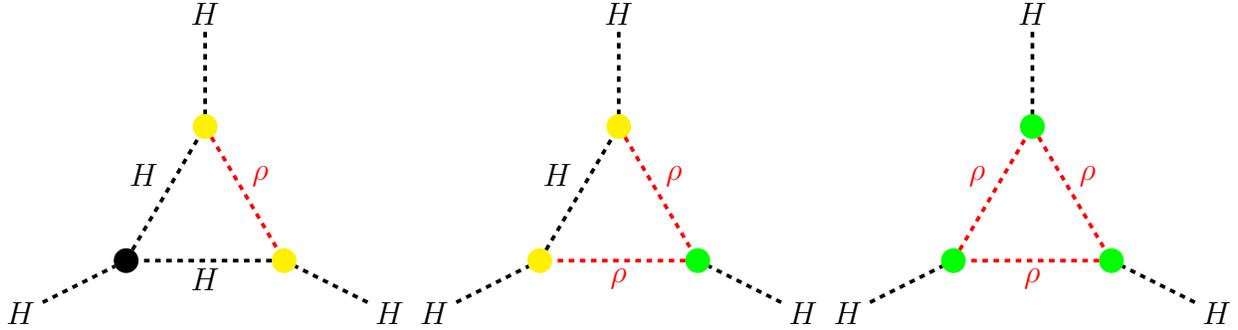
% http://texdoc.net/texmf-dist/doc/latex/tikz-feynhand/tikz-feynhand.userguide.pdf

\noindent The portal term of the Higgs potential contains the trilinear couplings for $HH\rho$ and $H\rho\rho$ vertices. The vertex factors for $HH\rho$ and $H\rho\rho$ vertices are introduced in Fig. \ref{fig:feynman}. The one-loop diagrams contributing to SM triple Higgs coupling are in Fig. \ref{fig:loop}. We denote the SM tree-level triple Higgs coupling as $\lambda_{HHH}$. The correction is gained by adding all the triangle diagrams (taking into account the symmetry factors):
\bea 
\Delta \lambda_{HHH} &=& 3\cdot 2 \cdot \lambda_{HHH} \lambda_{H\sigma}^2 v_\sigma^2 I(m_H,m_H,m_\rho; p,q)+ 3\cdot 2\cdot \lambda_{H\sigma}^3 vv_\sigma^2 I(m_H,m_\rho,m_\rho;p,q)\nonumber\\
&&+\;6\cdot \lambda_{H\sigma}^3 v^3 I(m_\rho,m_\rho,m_\rho;p,q).
\eea 
Here $p$ and $q$ are the external momenta and the loop integral is defined as
\be 
I(m_A,m_B,m_C;p,q) = \int \frac{d^4k}{(2\pi)^4}\frac{1}{(k^2-m_A^2)((k-p)^2-m_B^2)((k+q)^2-m_C^2)}.
\ee 
The process $H \rightarrow HH$ is disallowed for on-shell external momenta, so at least one of them must be off-shell. The first diagram is dominant due to heaviness of the $\rho$ scalar. Therefore, we may ignore the subleading contributions of diagrams involving two or more $\rho$ propagators. Integrating out the heavy scalar and calculating the integral,
\be 
\begin{split} 
\Delta \lambda_{HHH} &= -\lambda_{HHH}\frac{3\delta}{16\pi^2}\left( 2+ \ln \frac{\mu^2}{ m_H^2}-z\ln \frac{z+ 1}{z- 1}\right) 
\end{split} 
\ee 
where $ z \equiv \sqrt{1+(4m_H^2/q^2)}$ and $\mu = M_{Pl} = 1.22\times 10^{19}$ GeV is the regularization scale (in our analysis, chosen to be the Planck scale). We have used the modified minimal subtraction scheme ($\overline{\text{MS}}$), where the terms $\ln 4\pi$ and Euler-Mascheroni constant $\gamma_E \approx 0.57722$ emerging in the calculation are absorbed to the regularization scale $\mu$. Note that the correction is dependent on the Higgs off-shell momentum $q \equiv q^*$, which we assume to be at $\mathcal{O}(1)$ TeV at the LHC and HL-LHC. It is especially interesting to see that at the leading order, the triple Higgs coupling correction is proportional to the threshould corrections. This intimate connection forbids a too large correction. In fact, the bound from vacuum stability turns out to constrain the triple Higgs coupling correction to $\lesssim 25\%$, as we shall see in Section \ref{results}.

It should be noted that loop corrections contributing to the final to-be-observed value are included in the SM. Indeed, experiments are measuring $\lambda_{HHH}^\text{SM} = \lambda_{HHH}^\text{SM(tree)} + \lambda_{HHH}^\text{SM(1-loop)}(q^*) + \ldots$, where the SM one-loop correction depends on the Higgs off-shell momentum. At the $\mathcal{O}(1)$ TeV scale we are considering, the SM 1-loop correction amounts to approximately $-7\%$ \cite{Baglio:2016ijw}.

%%%%%%%%%%%%%%%%%%%%%%

\subsection{Light neutrino masses}

\noindent Neutrino sector of SMASH is able to generate correct neutrino masses and observed baryon asymmetry of the universe with suitable benchmarks. The relevant Yukawa terms for neutrinos in the model are

\be 
\mathcal{L}_\nu^Y = -\frac{1}{2}Y_n^{ij}\sigma N_iN_j - Y_\nu^{ij}L_i\varepsilon HN_j.
\ee 
We take a simplified approach: Dirac and Majorana Yukawa matrices ($Y_\nu$ and $Y_n$, respectively) are assumed to be diagonal.
\be 
Y_\nu = \left( \begin{array}{ccc}
y_1 & 0 & 0 \\ 0 & y_2 & 0 \\ 0 & 0 & y_3
\end{array}\right) , \quad Y_n = \left( \begin{array}{ccc}
Y_1 & 0 & 0 \\ 0 & Y_2 & 0 \\ 0 & 0 & Y_3
\end{array}\right) .
\ee 
To generate baryonic asymmetry of the universe, SMASH utilizes thermal leptogenesis scenario \cite{Fukugita86}, which generates lepton asymmetry in the early universe and leads to baryon asymmetry. In the scenario, heavy neutrinos require a sufficient mass hierarchy \cite{Buchmuller:2002rq,Davidson:2002qv} and one or more Yukawa couplings must have complex CP phase factors. We assume the CP phases are $\mathcal{O}(1)$ radians to near-maximize the CP asymmetry \cite{Buch05,Buch13,Buch14}
\be 
\varepsilon_\text{CP} = \frac{\Gamma(N_1 \rightarrow H + \ell_L)-\Gamma(N_1 \rightarrow H^\dagger + \ell_L^\dagger)}{\Gamma(N_1 \rightarrow H + \ell_L)+\Gamma(N_1 \rightarrow H^\dagger + \ell_L^\dagger)} \lesssim \frac{3M_1m_3}{16\pi v^2}.
\ee 
The largest value is obtained if the CP violation is maximal. Large asymmetry is needed for producing matter-antimatter asymmetry in the unverse. Following \cite{Smash1}, we set heavy neutrino mass hierarchy $M_3 = M_2 = 3M_1$, corresponding to $Y_3 = Y_2 = 3Y_1$. These choices give the full $6 \times 6$ neutrino mass matrix
\be 
M_\nu = \left( \begin{array}{cc}
\textbf{0}_{3 \times 3} & m_D \\ m_D^T & M_M,
\end{array}\right),
\ee
which is in block form, and contains five free parameters: $v_\sigma, y_1, y_2, y_3$ and $Y_1$. Here $m_D = Y_\nu v/\sqrt{2}$ is the Dirac mass term and $M_M = Y_nv_\sigma/\sqrt{2}$ is the Majorana mass term. Light neutrino masses are then generated via well-known Type I seesaw mechanism \cite{Fritzsch75,Minkowski:1977sc,GellMann:1980vs,Yanagida:1979as,Mohapatra:1979ia,Mohapatra:1980yp,Schechter:1980gr,Magg:1980ut,Glashow:1979nm,Lazarides:1980rn,Gelmini:1980re}, by block diagonalizing the full neutrino mass matrix $M_\nu$.

It is possible to obtain light neutrino masses consistent with experimental constraints from atmospheric and solar mass splittings $\Delta m_{32}^2$ and $\Delta m_{21}^2$ and cosmological constraint $m_1+m_2+m_3 < 0.12$ eV (corresponding to $m_1 \lesssim 0.03$ (0.055) eV with normal (inverse) neutrino mass ordering), assuming the standard $\Lambda$CDM cosmological model. 

The light neutrino mass matrix $m_\nu =-(v^2/(\sqrt{2}v_\sigma))Y_\nu Y_n^{-1} Y_\nu^T$ is (after removing the irrelevant sign via field redefinition)
\be 
m_\nu = C \left( \begin{array}{ccc}
y_1^2/Y_1 & 0 & 0 \\ 0 &y_2^2/Y_2 & 0 \\ 0 & 0 & y_3^2/Y_3
\end{array}\right) = \left( \begin{array}{ccc}
m_1 & 0 & 0 \\ 0 &\sqrt{m_1^2 + \Delta m_{31}^2} & 0 \\ 0 & 0 & \sqrt{m_1^2 + \Delta m_{21}^2 + \Delta m_{32}^2}
\end{array}\right) ,
\ee 
where we have denoted $C=v^2/(\sqrt{2}v_\sigma)$ and assumed normal mass ordering. This gives the neutrino masses $m_i = Cy_i^2/Y_i$. We do not know the absolute masses, but the mass squared differences have been measured by various neutrino oscillation experiments \cite{Esteban:2018azc}. Nevertheless, their values provide two constraints, leaving three free parameters. However, the heavy neutrino Yukawa couplings $Y_i$ must be no larger than $\mathcal{O}(10^{-3})$ to avoid vacuum instability \cite{Smash2}.

In addition, an order-of-magnitude estimate of generated matter-antimatter asymmetry (baryon-to-photon ratio) is directly proportional to the CP asymmetry:
\be
\eta \equiv \frac{n_B}{n_\gamma} = \mathcal{O}\left(10^{-2}\right)\varepsilon_\text{CP} \kappa,
\ee 
where $\kappa \sim 0.01 - 0.1$ is an efficiency factor. A more precise value of $\kappa$ can be determined by solving the Boltzmann equations, which is outside the scope of this study. We arrive at
\be \label{eta}
\eta = \mathcal{O}\left(10^{-10}\right) \times \frac{v_\sigma}{10^8 \text{ GeV}} \times \frac{Y_1}{10^{-2}}\times \frac{\kappa}{0.1},
\ee 
which in principle can be consistent with the observed $\eta$. We will provide suitable benchmark points in the next Section.

%%%%%%%%%%%%%%%%%%%%%%

\section{\label{methods}Methods}

\begin{table}[]
\begin{center}
{\renewcommand{\arraystretch}{1.5}\setlength{\tabcolsep}{0.5em}   
\begin{tabular}{|c|c|c|c|c|}\hline 
\rule{0pt}{3ex}\textbf{Benchmarks} & \textbf{BP1} & \textbf{BP2} & \textbf{BP3} & \textbf{BP4}\\\hline 
\rule{0pt}{3ex}$Y^\nu_{11}$ & $5.073\times 10^{-5}$ & $3.481\times 10^{-5}$ & $1.373\times 10^{-7}$ & $4.446\times 10^{-4}$ \\\hline 
\rule{0pt}{3ex}$Y^\nu_{22}$ & $9.059\times 10^{-5}$ & $2.225\times 10^{-4}$ & $9.376\times 10^{-6}$ & $9.080\times 10^{-4}$ \\\hline 
\rule{0pt}{3ex}$Y^\nu_{33}$ & $1.341\times 10^{-4}$ & $5.377\times 10^{-4}$ & $2.266\times 10^{-5}$ & $1.844\times 10^{-3}$ \\\hline 
\rule{0pt}{3ex}$Y^N_{11}$ & $4.691\times 10^{-3}$ & $8.226\times 10^{-3}$ & $1.461\times 10^{-3}$ & $9.521\times 10^{-3}$ \\\hline 
\rule{0pt}{3ex}$Y_Q$ & $ 10^{-3}$ & $ 10^{-3}$ & $ 10^{-3}$ & $ 10^{-3}$ \\\hline 
\rule{0pt}{3ex}$v_\sigma$ (GeV) & $10^9 $ & $10^{10} $ & $10^8 $ & $10^{11} $ \\\hline 
\rule{0pt}{3ex}$\lambda_{\sigma}$ & $1.5 \times 10^{-10}$ & $10^{-9}$ & $10^{-10} $ & $1.6\times 10^{-9} $ \\\hline 
\rule{0pt}{3ex}$\lambda_{H\sigma}$ & $-1.6 \times 10^{-6}$ & $-1.2\times10^{-5}$ & $-1.5\times 10^{-6} $ & $-5\times 10^{-6} $ \\\hline 
\rule{0pt}{3ex}$q$ & $-1/3$ & $-1/3$ & $-1/3$ & $-1/3$ \\\hline 
\rule{0pt}{3ex}$\kappa$ & $0.1$ & $0.1$ & $0.1$ & $0.1$ \\\hline 
\rule{0pt}{3ex} $q^*$ (GeV) & 1000 & 1000 & 1000 & 1000 \\\hline 
\end{tabular}}
\end{center}
\caption{\label{BM}Used benchmark points (BP) in our analysis. Note that we assume specific texture to right-handed neutrino Yukawa matrix $Y^n$. $q$ refers to the charge of the extra quark-like electroweak singlet particle. $\kappa$ is the efficiency factor used in calculation of matter-antimatter asymmetry. $q^*$ is the value of off-shell momentum used in the calculation of triple Higgs coupling correction.}
\end{table}

\begin{table}[]
\begin{center}
{\renewcommand{\arraystretch}{1.5}\setlength{\tabcolsep}{0.5em}
\begin{tabular}{|c|c|c|c|c|c|c|c|c|}\hline
\textbf{Parameter} & $m_t^{\overline{\text{MS}}}(m_t)$ & $m_b$ & $m_H$ & $m_\tau$ & $v$ & $g_1$ & $g_2$ & $g_3$\\ \hline 
\textbf{Value} & 164.0 & 4.18 & 125.18 & 1.777 & 246.22 & 0.357 & 0.652 & 1.221 \\ \hline 
\end{tabular}}
\end{center}
\caption{\label{SM}Used SM inputs in our analysis, at $\mu = m_Z = 91.18$ GeV, with the exception of top mass, which is evaluated at $\mu=m_t$. The masses and vacuum expectation value are in GeV units.}
\end{table}

We generate the suitable benchmark points demonstrating different physics aspects of the model in the neutrino sector by fitting in the known neutrino mass squared differences $\Delta m_{ij}^2$, assuming normal mass ordering $(m_1 < m_2 < m_3)$. This leaves three free neutrino parameters, values of which we generate by logarithmically distributed random sampling. These are the candidates for benchmark points. We then require that the candidate points are consistent with the bound for the sum of light neutrino masses. The next step is to choose the suitable values of other unknown parameters, using the stability of the vacuum as a requirement.

The authors of \cite{Smash1} have generated the corrections to two-loop $\beta$ functions of SMASH. We solve numerically the full two-loop 14 coupled renormalization group differential equations with SMASH corrections with respect to Yukawa ($Y^u, Y^d, Y^e, Y^\nu, Y^N, Y^Q$), gauge ($g_1,g_2,g_3$) and scalar couplings ($\mu_H^2, \mu_S^2, \lambda_H, \lambda_\sigma, \lambda_{H\sigma}$), ignoring the light SM degrees of freedom, from $M_Z$ to Planck scale. We assume Yukawa matrices are in diagonal basis. We use $\overline{\text{MS}}$ scheme for the running of the RGE's. Since the top quark $\overline{\text{MS}}$ mass is different from its pole mass, the difference is taken into account via the relation \cite{Jegerlehner:2012kn}
\be 
m_t^\text{pole}\approx m_t^{\overline{\text{MS}}}\left(1 + 0.4244\alpha_3 + 0.8345\alpha_3^2 + 2.375\alpha_3^3 + 8.615\alpha_3^4\right),
\ee 
where $\alpha_3 \equiv g_3^2/4\pi \approx 0.1085$ at $\mu = m_Z$. We define the Higgs quadratic coupling as $\mu_H = m_H/\sqrt{2}$ and quartic coupling as $\lambda_H = m_H^2/2v^2$.

We use MATLAB R2019's \texttt{ode45}-solver. See Table \ref{BM} for used SMASH benchmark points, and Table \ref{SM} for our SM input. Our scale convention is $t \equiv \log_{10} \mu/$ GeV.

In some papers, the running of SM parameters ($Y^t, Y^b, Y^\tau, g_1, g_2, g_3, \mu_H^2, \lambda_H$) obeys the SM RGE's without corrections from a more effective theory until some intermediate scale $\Lambda_\text{BSM}$ \cite{Elias-Miro2012}, after which the SM parameters gain threshold correction (where it is relevant) and the running of all SM parameters follows the new RGE's from that point onwards. We choose to utilize this approach while acknowledging an alternative approach, where the threshold correction is applied at the beginning ($\mu = m_Z$) \cite{Lebedev2012}, and both approaches give the almost same results. As previously stated, SM Higgs quadratic and quartic couplings will gain the threshold correction.

Our aim is to find suitable benchmark points, which
\begin{itemize}
\item allow the quartic and Yukawa couplings of the theory to remain positive and perturbative up to Planck scale,
\item utilize threshold correction mechanism to $\lambda_H$ via $\delta \approx 0.01 - 0.1$,
%\item avoid the overproduction of dark radiation via the cosmic axion background (requiring $\lambda_{H\sigma} < 0$),
\item produce a significant contribution matter-antimatter asymmetry via leptogenesis (requiring hierarchy between the heavy neutrinos), and
\item produce a large enough correction to triple Higgs coupling $\lambda_{HHH}$ to be detected in the future by HL-LHC or FCC-hh.
\end{itemize}

\begin{figure}
\centering
\includegraphics[width=0.496\linewidth]{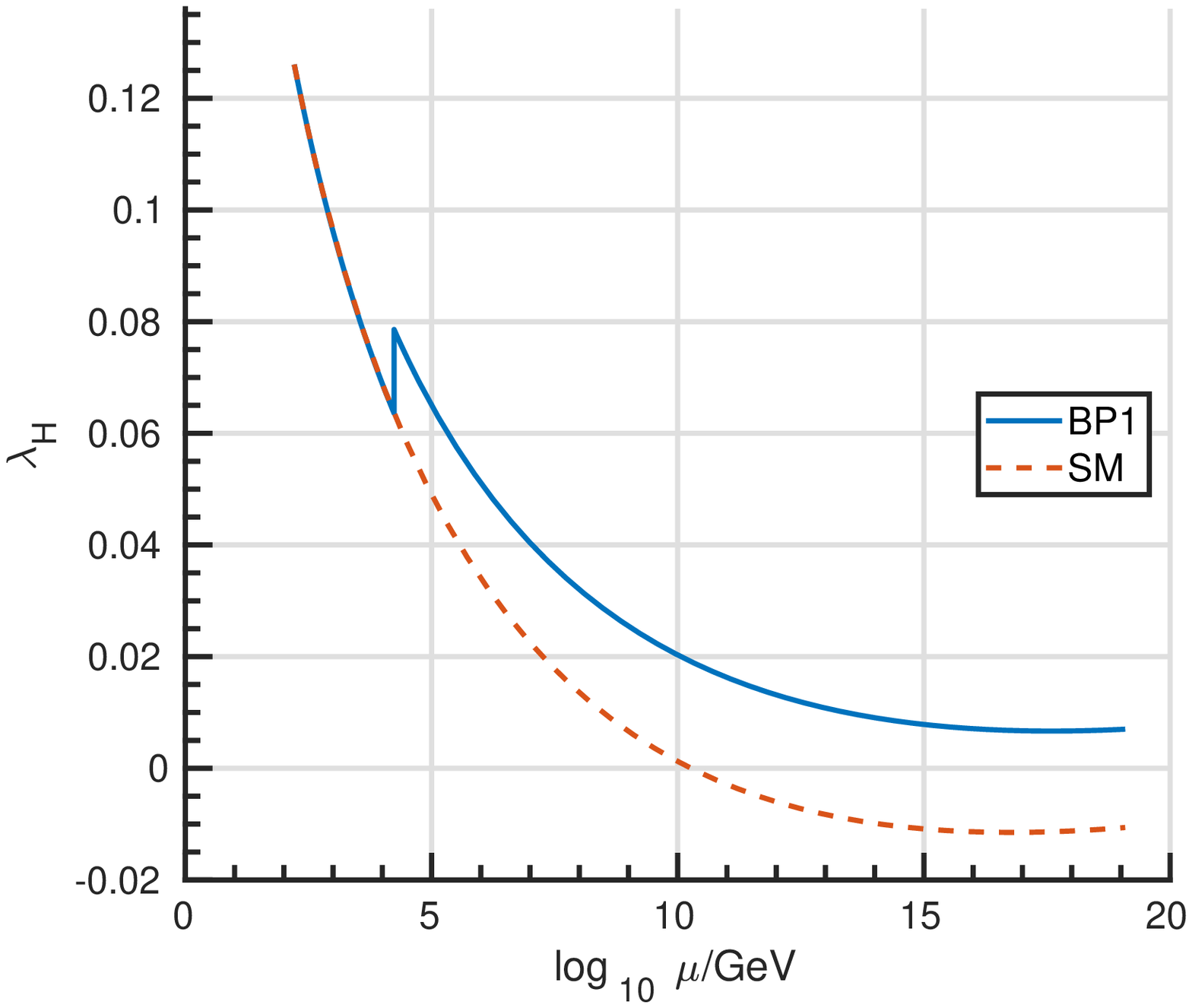}
\includegraphics[width=0.496\linewidth]{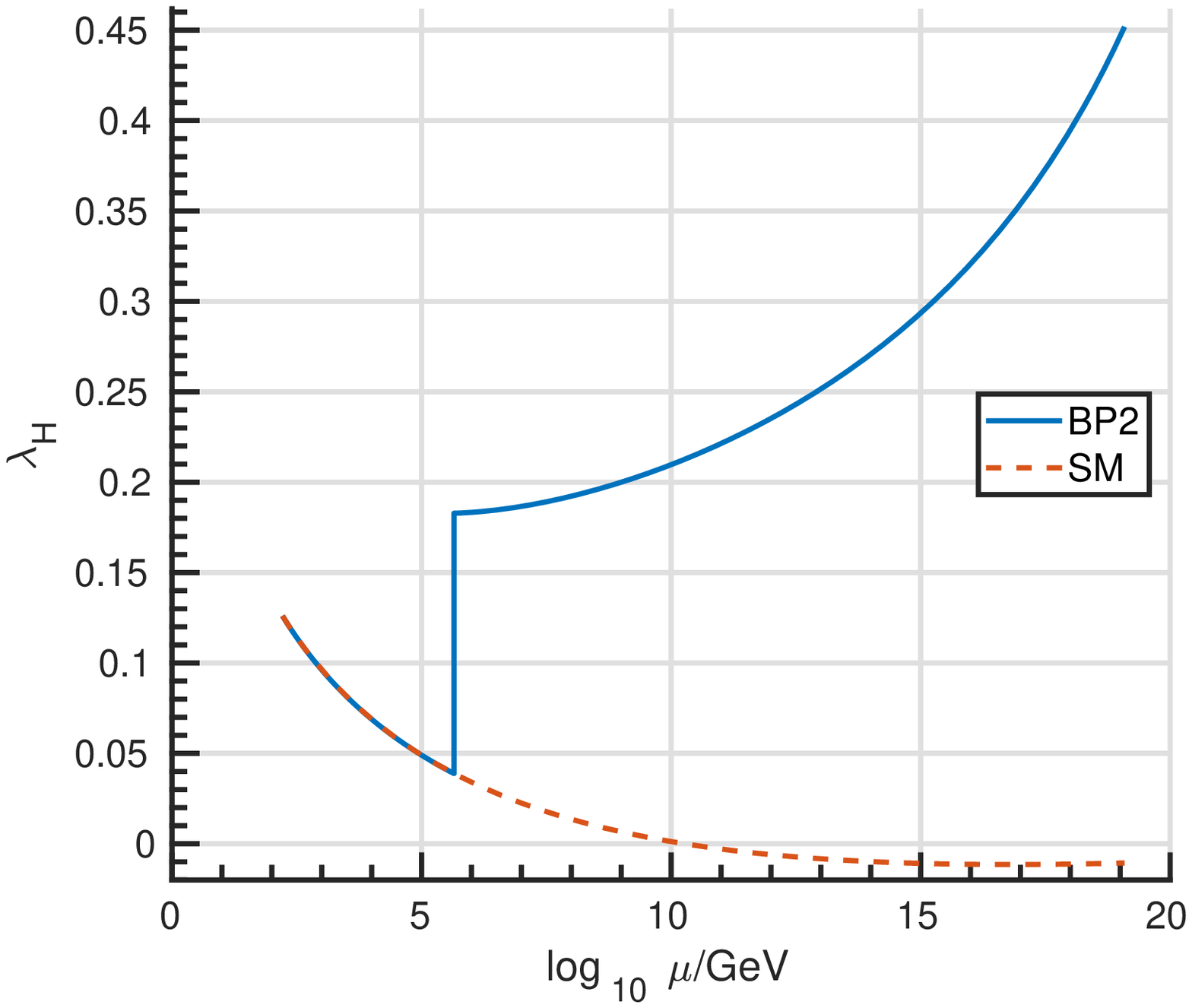}\\
\includegraphics[width=0.496\linewidth]{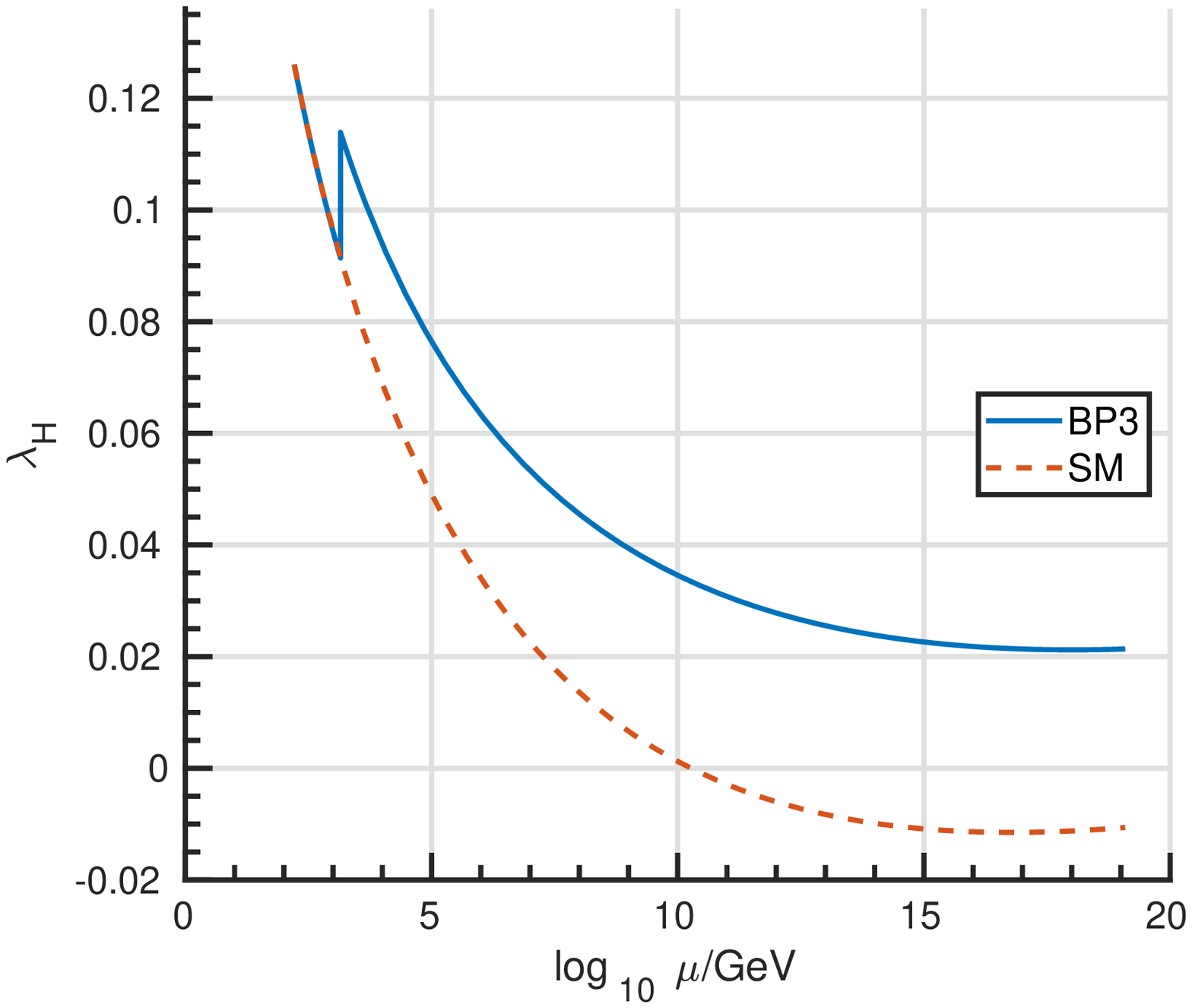}
\includegraphics[width=0.496\linewidth]{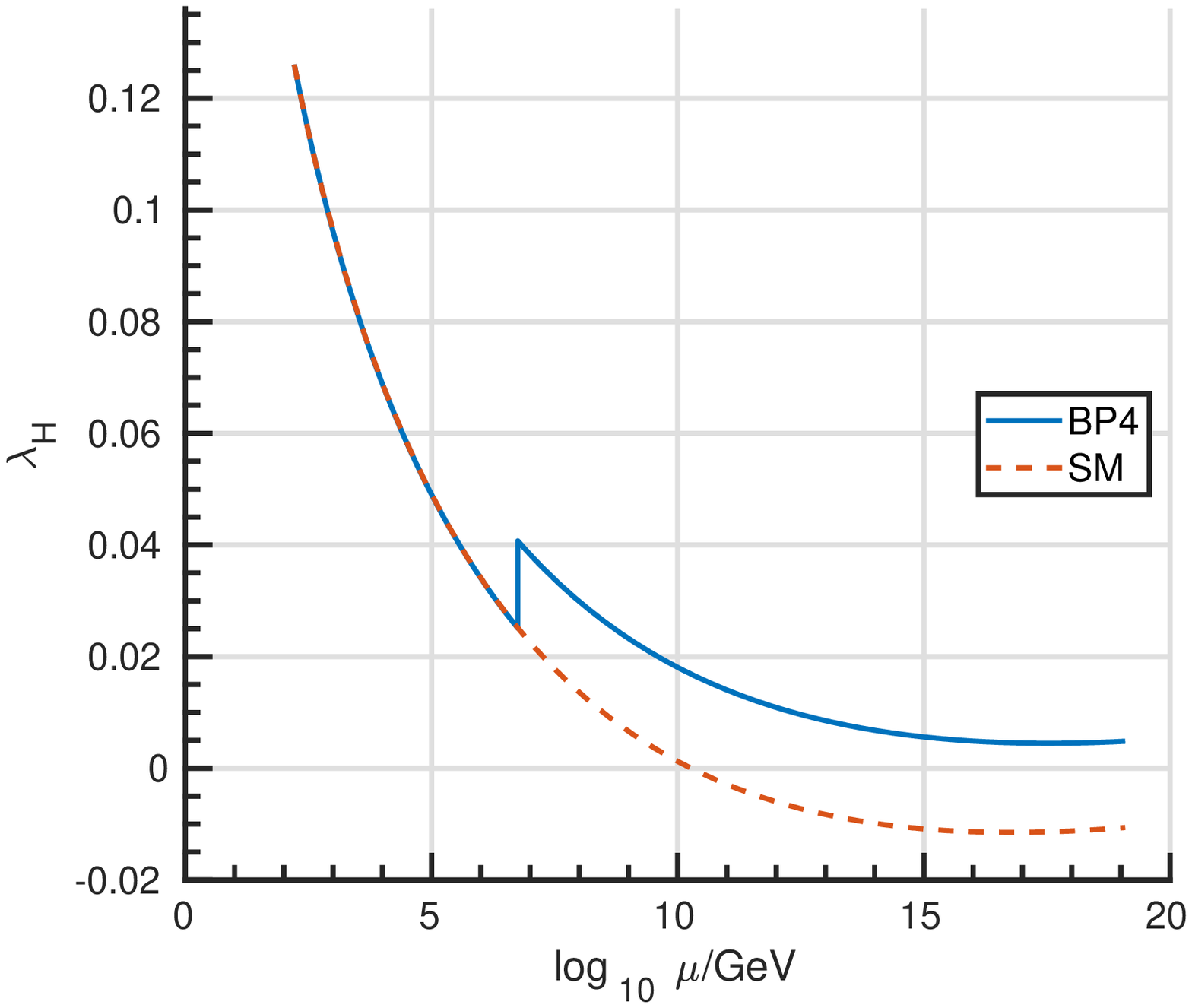}
\caption{Running of SM Higgs quartic coupling in Standard Model (dashed line) and in SMASH with benchmark points \textbf{BP1}-\textbf{BP4} (solid line). Threshold correction is utilized at $m_\rho$.}
\label{fig:threshold}
\end{figure}

%%%%%%%%%%%%%%%%%%%%%%

\section{\label{results}Results}

%%%%%%%%%%%%%%%%%%%%%%

\subsection{Stability of vacuum}

\begin{figure}
\centering
\includegraphics[width=0.496\linewidth]{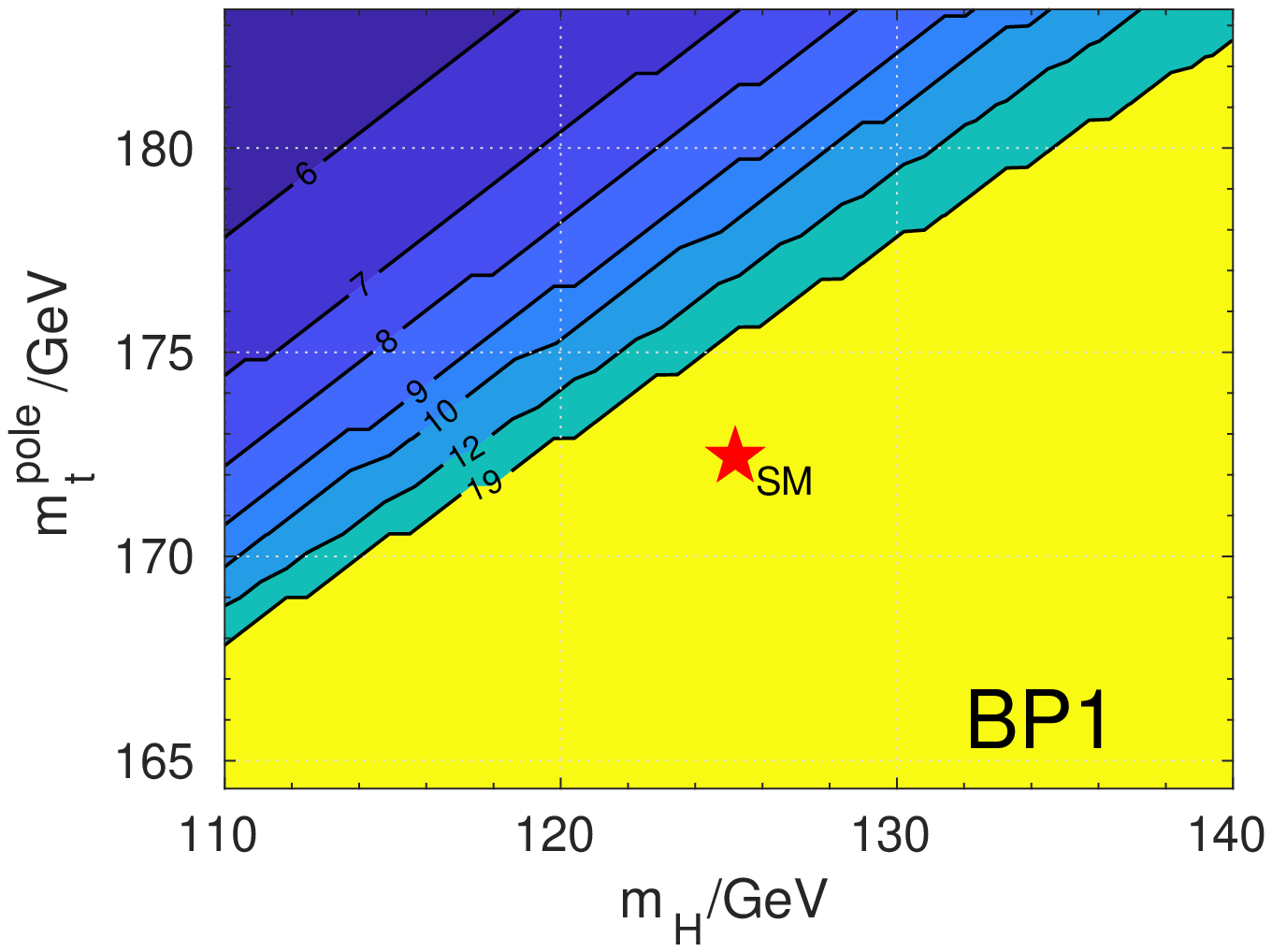}
\includegraphics[width=0.496\linewidth]{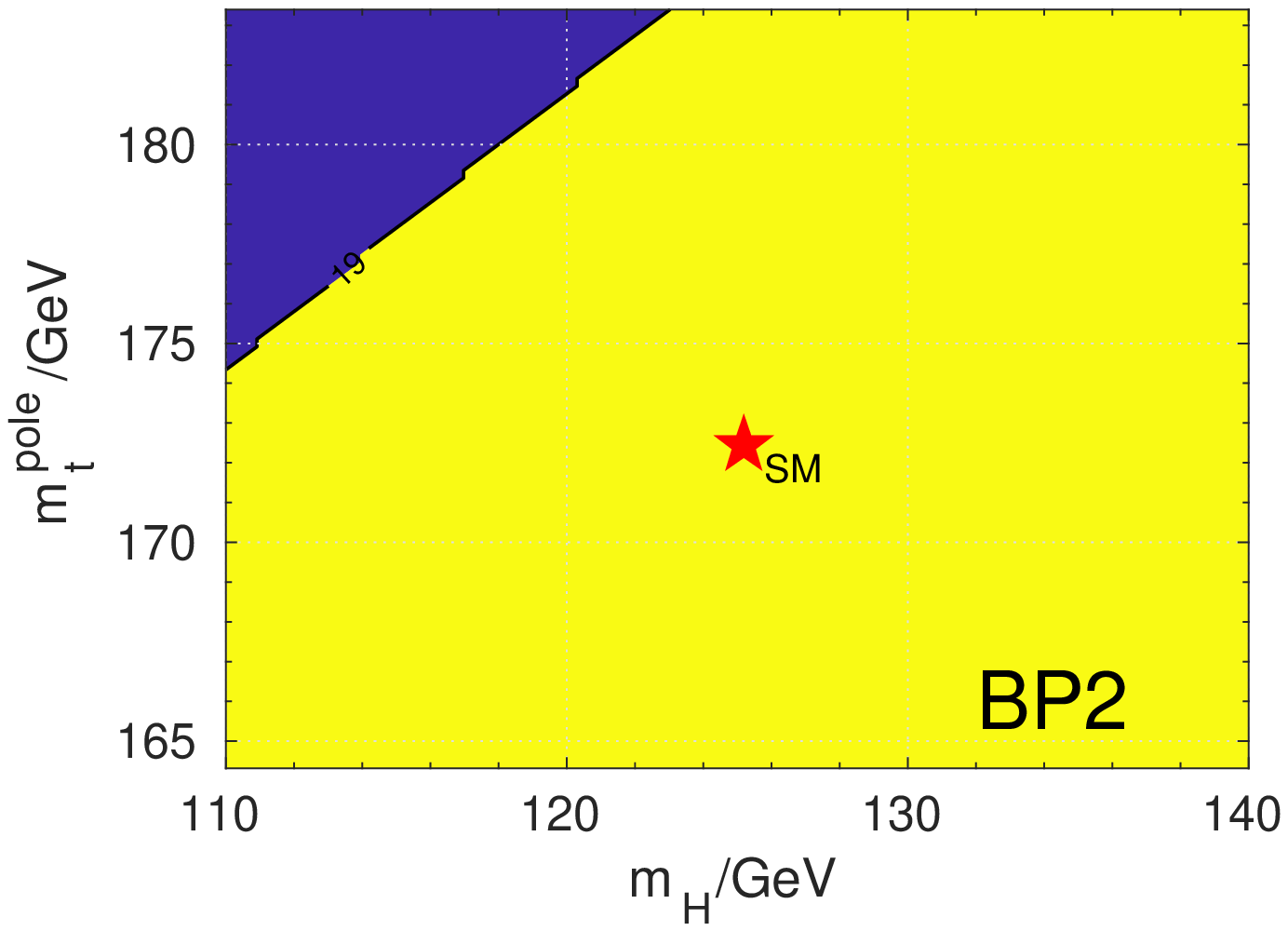}\\
\includegraphics[width=0.496\linewidth]{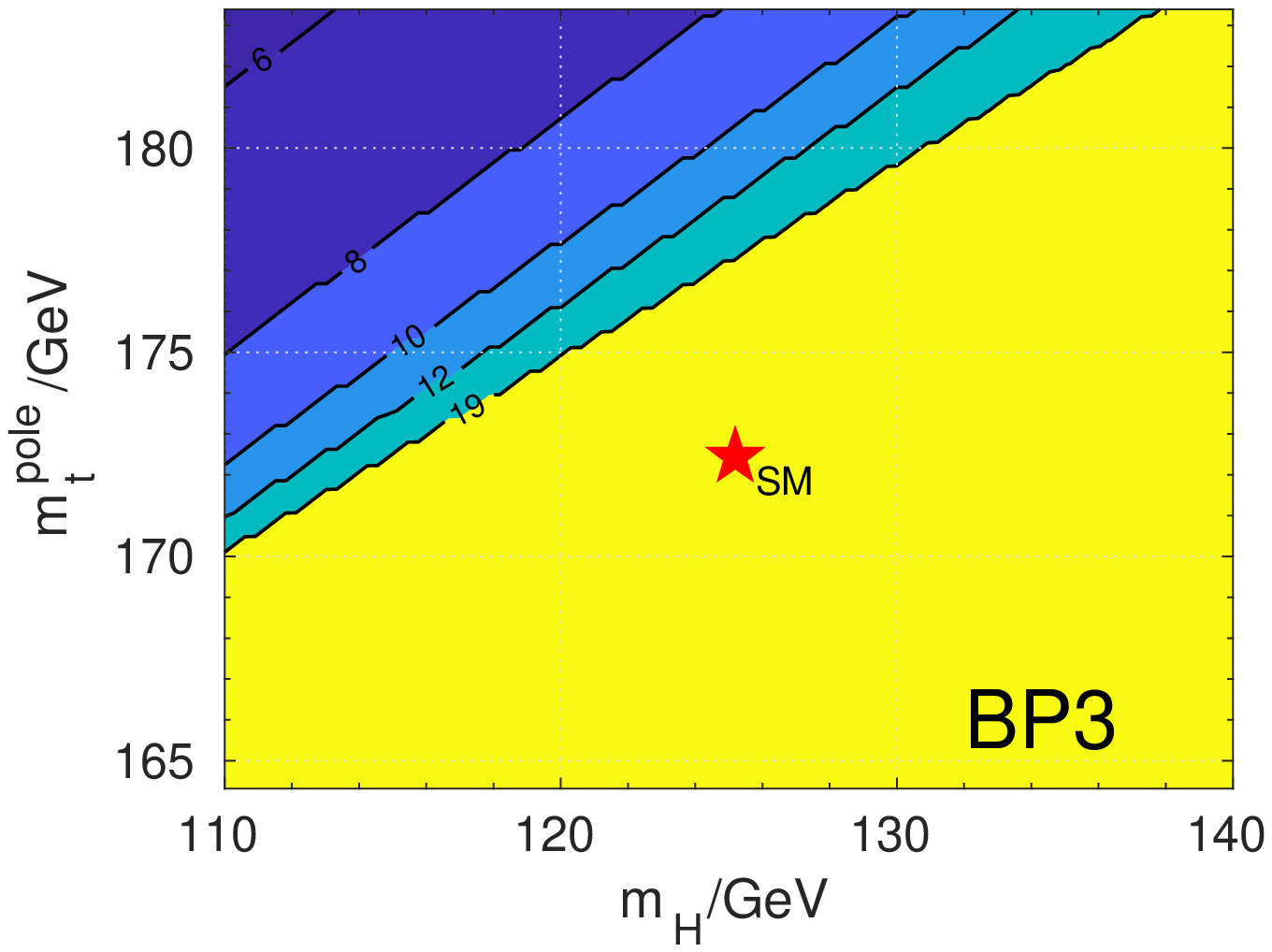}
\includegraphics[width=0.496\linewidth]{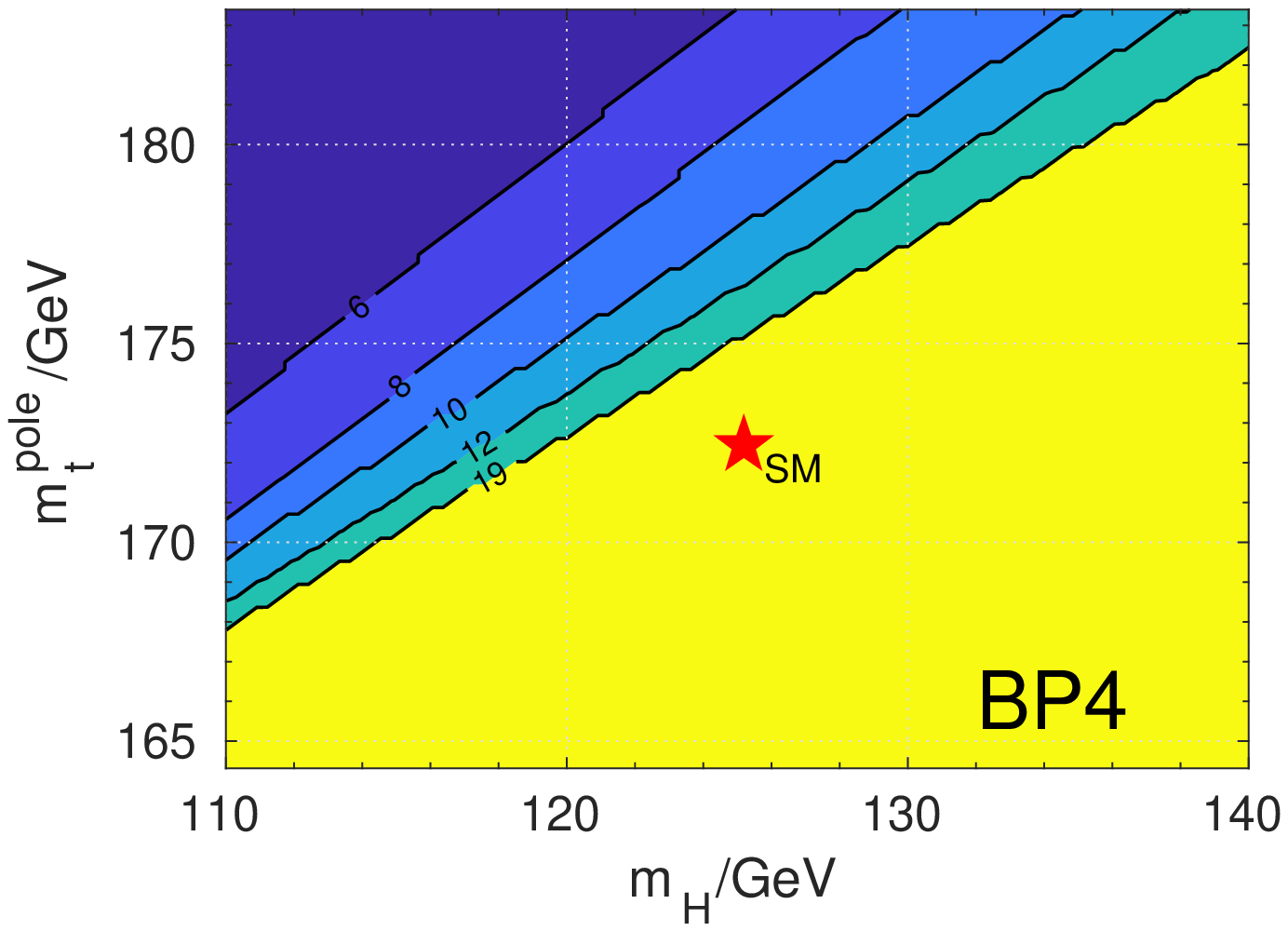}
\caption{\label{mh-mt}Vacuum stability of SMASH in $(m_H,m_t^\text{pole})$ plane with benchmark points \textbf{BP1}-\textbf{BP4}. The red star corresponds to the SM best-fit value. The height and width of the star corresponds to the present uncertainties. Vacuum is stable in yellow region. The contour numbers $n$ correspond to the vacuum instability scale $10^n$ GeV.} % In the cyan region one or more quartic couplings exceed unity below the Planck scale.
\label{fig:bp-mhmt}
\end{figure}

We have plotted how the running of the SM quartic coupling $\lambda_H$ changes with each benchmark point in Fig. \ref{fig:threshold}. Note that all the threshold corrections are utilized well before the SM instability scale $\Lambda_\text{IS}$.

We numerically scanned over the parameter space $m_t^\text{pole} \in [164, 182]$ GeV and $m_H \in [110,140]$ GeV to analyze vacuum stability in four different benchmark points \textbf{BP1}-\textbf{BP4}. Our result for the chosen benchmarks is in Fig. \ref{fig:bp-mhmt}, where the SM best fit is denoted by red star. Clearly the electroweak vacuum is stable with our benchmark points and it assigned to $m_t^\text{pole} \simeq 173.0 \pm 0.4$ GeV and $m_H \simeq 125.18 \pm 0.16$ GeV \cite{PDG18}. For every case, we investigated the running of the quartic couplings of the scalar potential. If either $\lambda_H$ or $\lambda_{\sigma}$ turn negative, we denote this point unstable. If any of the quartic couplings rise above $\sqrt{4\pi}$, we denote this point non-perturbative.

\begin{figure}
\begin{center}
\includegraphics[width=0.496\linewidth]{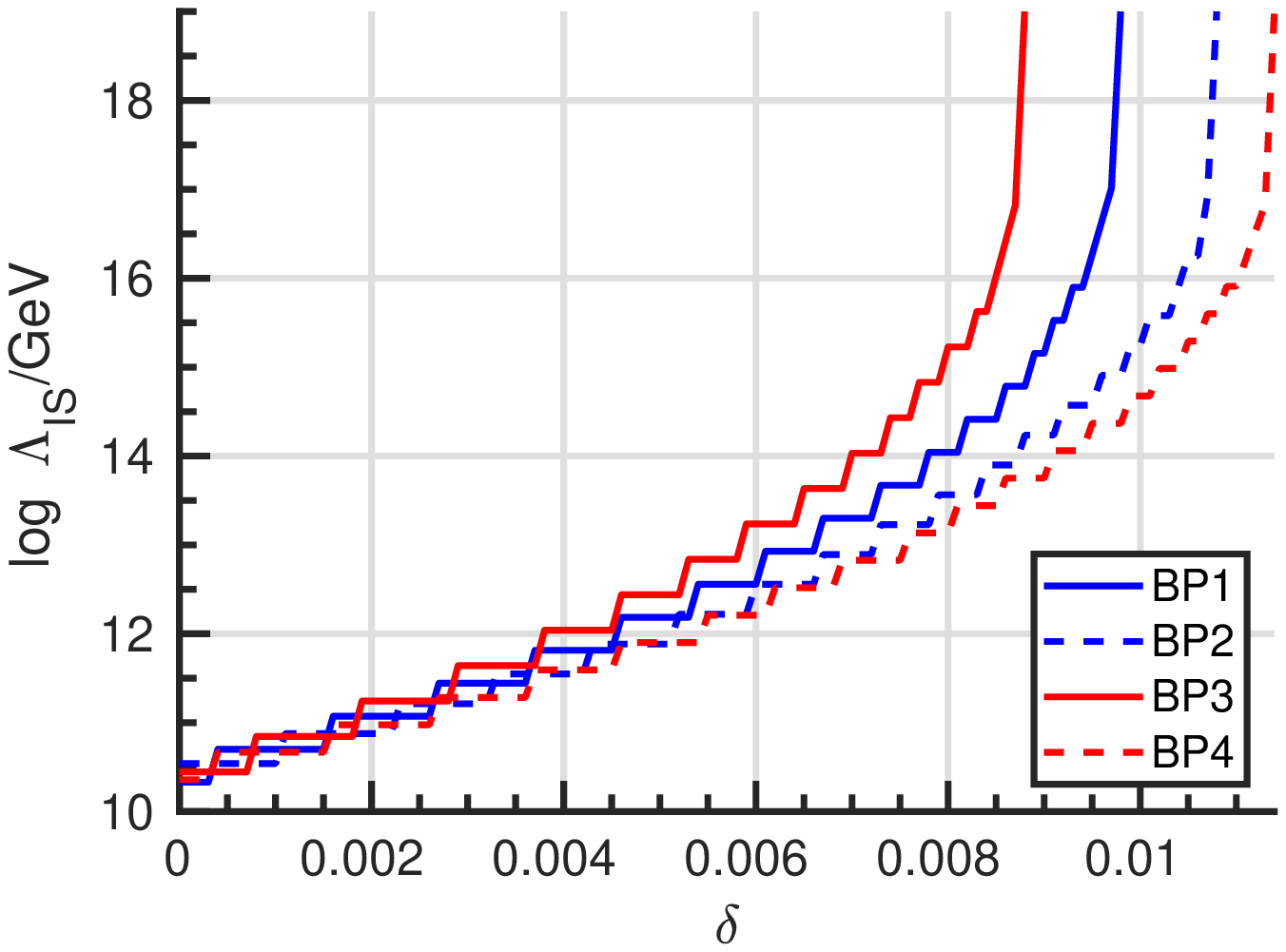}
\includegraphics[width=0.496\linewidth]{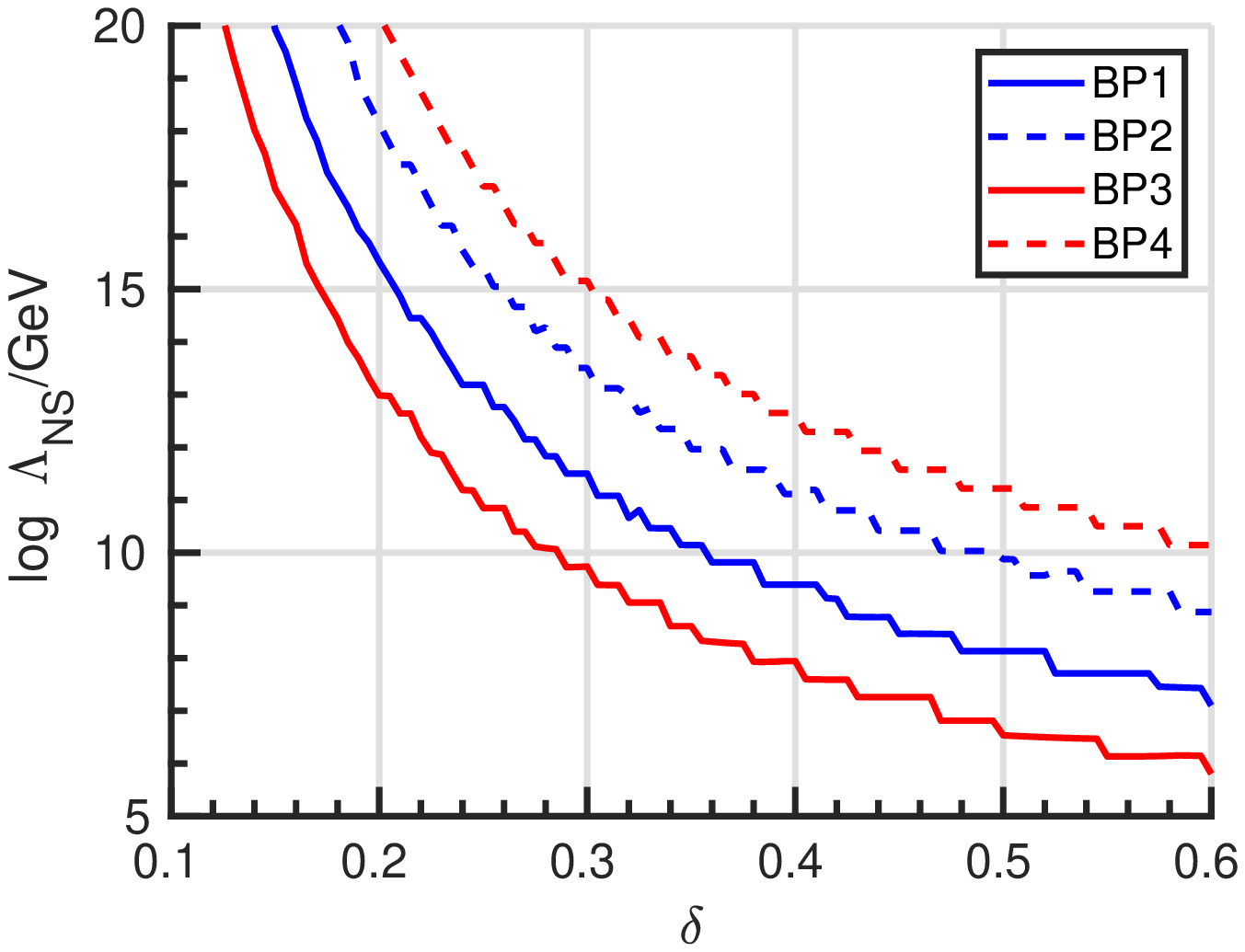}
\caption{\label{fig:instability-scale}The rise of instability scale (left) and the fall of nonperturbativity scale (right) as a function of threshold correction $\delta$, for \textbf{BP1}-\textbf{BP4}.}
\end{center}
\end{figure}

In \textbf{BP2}, we have chosen the new scalar parameters in such a way that the threshold correction is large, $\delta > 0.1$. This changes the behaviour of the running so that after the correction the $\lambda_H$ \textit{increases} in energy instead of decreasing, opposite to the coupling's running in pure SM scenario. A too-large threshold correction will have an undesired effect, lowering the nonperturbativity scale to energies lower than the Planck scale. These effects are visualized in Fig. \ref{fig:instability-scale}, where for each benchmark point kept $\lambda_\sigma$ at its designated value in Table \ref{BM}. Instead, we let the portal coupling $\lambda_{H\sigma}$ vary between 0 and $\sqrt{0.6\lambda_S}$. This demonstrates the small range of viable parameter space.

\begin{figure}[!h]
\begin{center}
\includegraphics[width=0.792\linewidth]{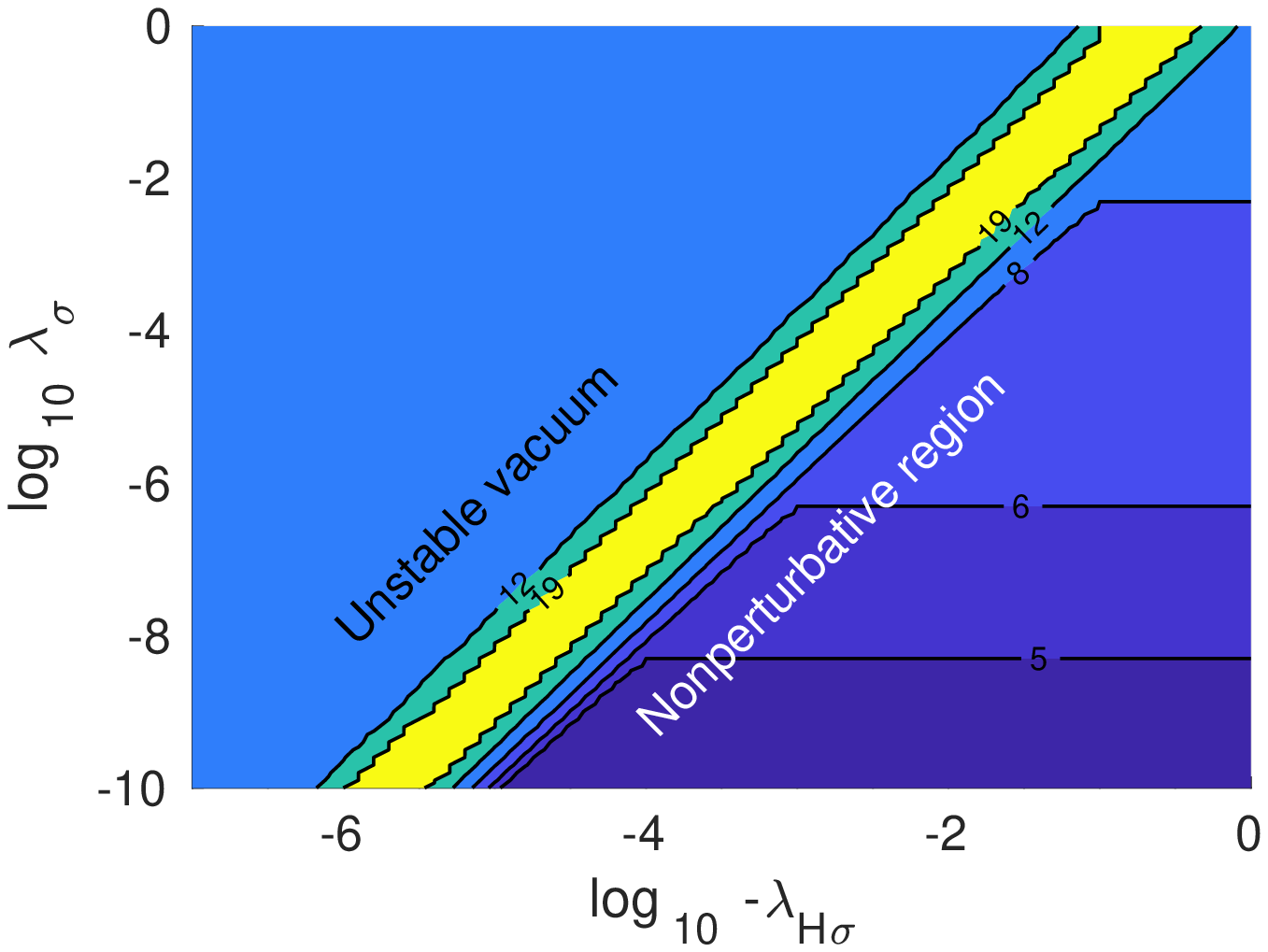}
\includegraphics[width=0.792\linewidth]{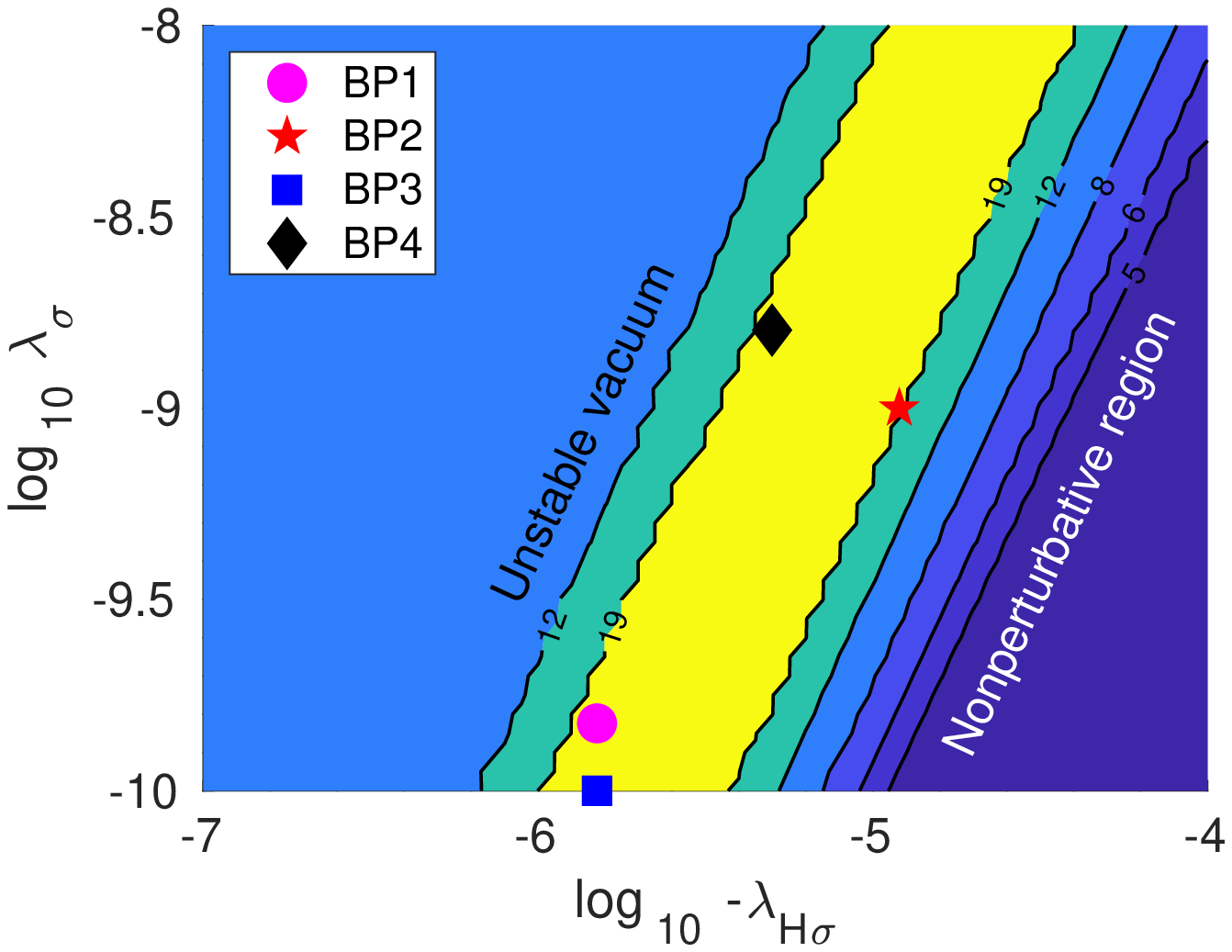}
\caption{\label{fig:lambda-2d} \textbf{Above:} Different regions in logarithmic $(-\lambda_{H\sigma},\lambda_\sigma)$ plane. The contour numbers $n$ above the yellow band correspond to vacuum instability scale $10^n$ GeV. Below the yellow band the contour numbers $m$ correspond to nonperturbativity scale $10^m$ GeV. The colour coding is interpreted as in Fig. \ref{fig:bp-mhmt}. For nonperturbative scale calculations, we have used \textbf{BP1}. \textbf{Below:} Zoomed-in detail of the figure above, showing in addition our chosen benchmarks.}
\end{center}
\end{figure}

Our next scan was over the new quartic couplings, $\log_{10} (-\lambda_{H\sigma} )\in [-7,0]$ and $\log_{10} \lambda_{\sigma} \in [-10,0]$. The scalar potential is stable and the couplings remain perturbative at only a narrow band, where $\delta \sim 0.01 - 0.1$, see Fig. \ref{fig:lambda-2d}. We chose \textbf{BP1}, \textbf{BP3} and \textbf{BP4} with small $\delta$, allowing the SM Higgs quartic coupling to decrease near zero at $\mu = M_{Pl}$. This reflects the placement of the benchmark points near the left side of the stability band. In contrast, we chose \textbf{BP2} with large $\delta$, placing it near the right side of the stability band, corresponding to large value of $\lambda_H$ at $\mu = M_{Pl}$.

\begin{figure}
\centering
\includegraphics[width=0.496\linewidth]{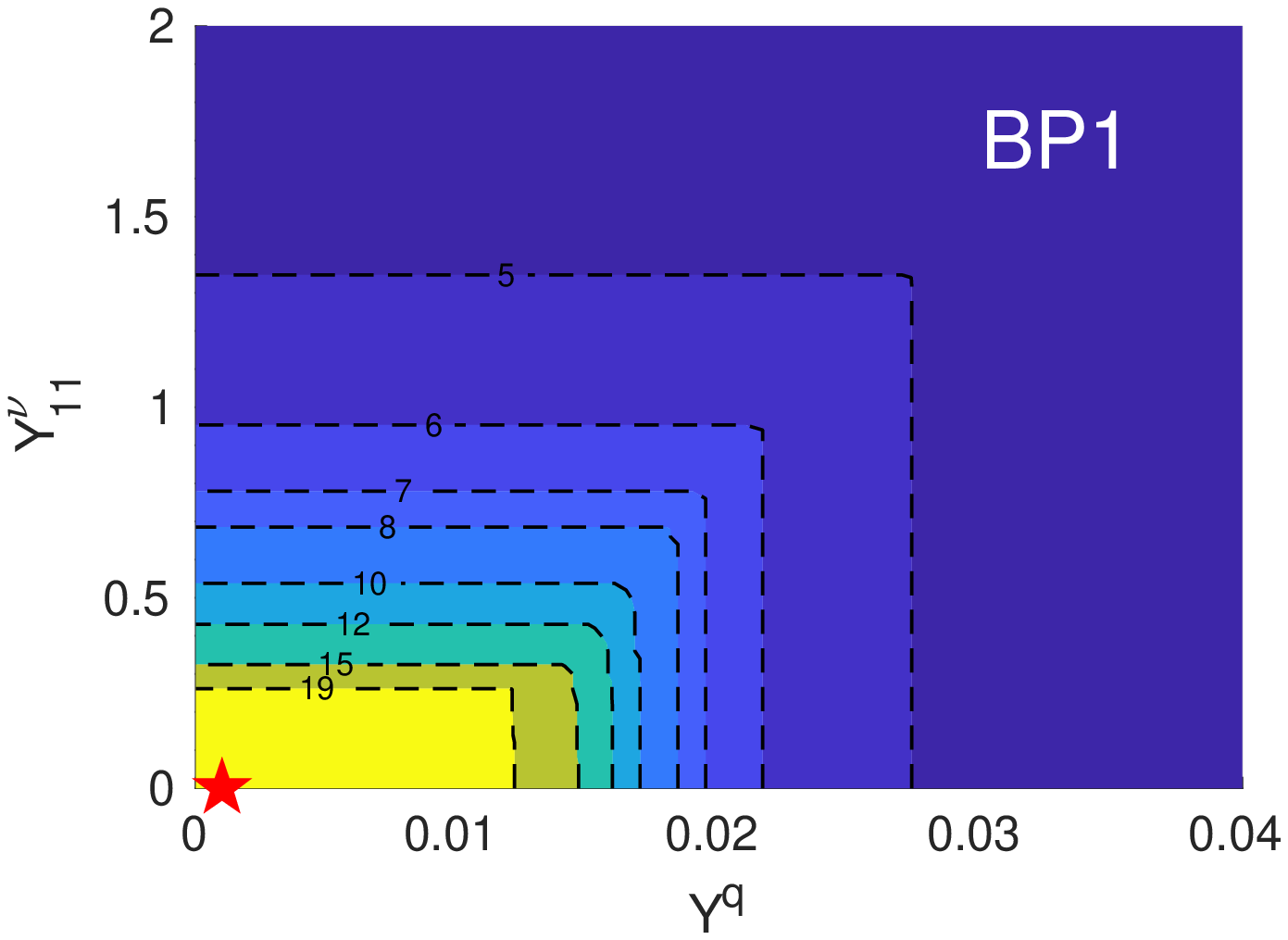}
\includegraphics[width=0.496\linewidth]{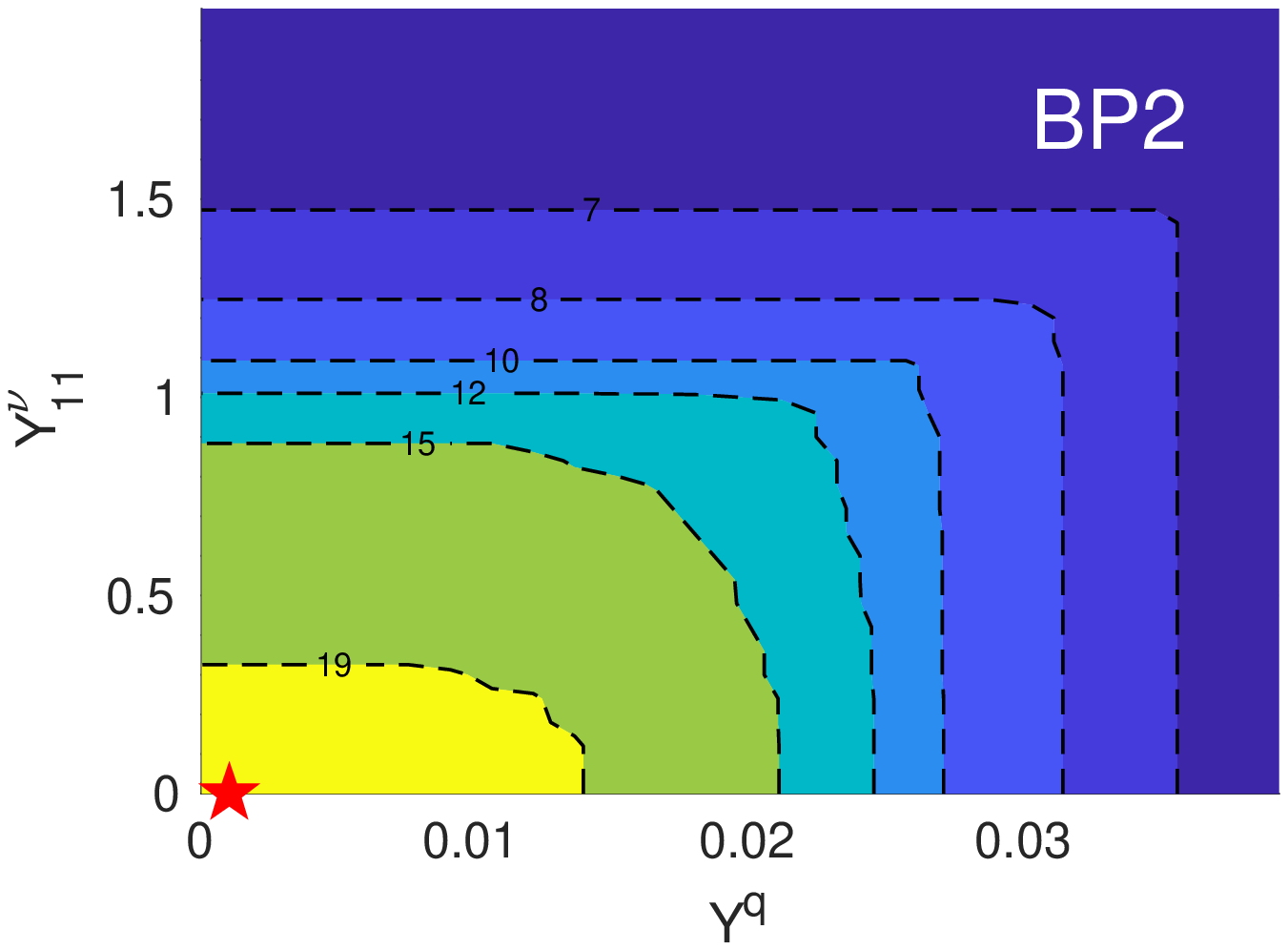}\\
\includegraphics[width=0.496\linewidth]{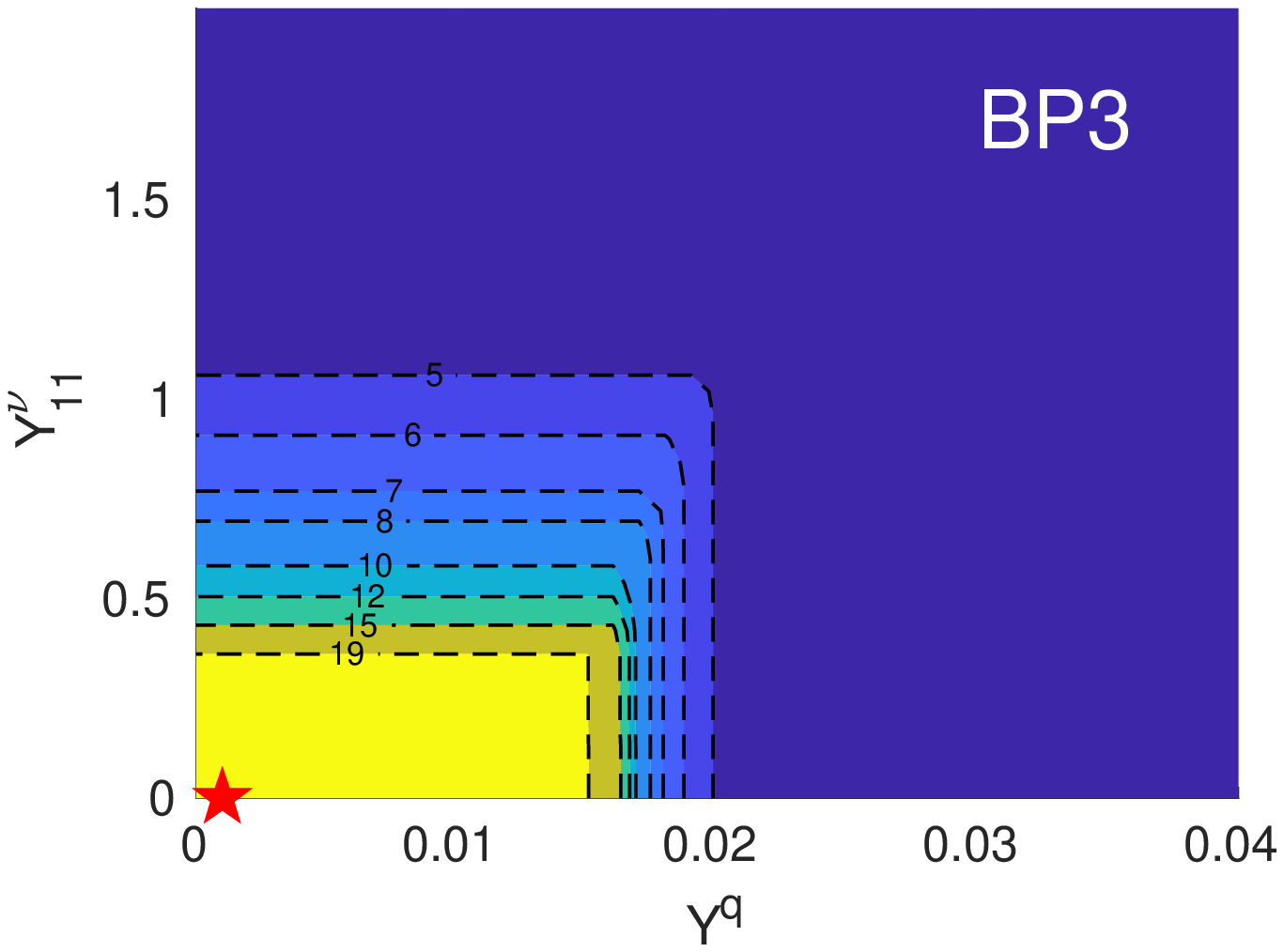}
\includegraphics[width=0.496\linewidth]{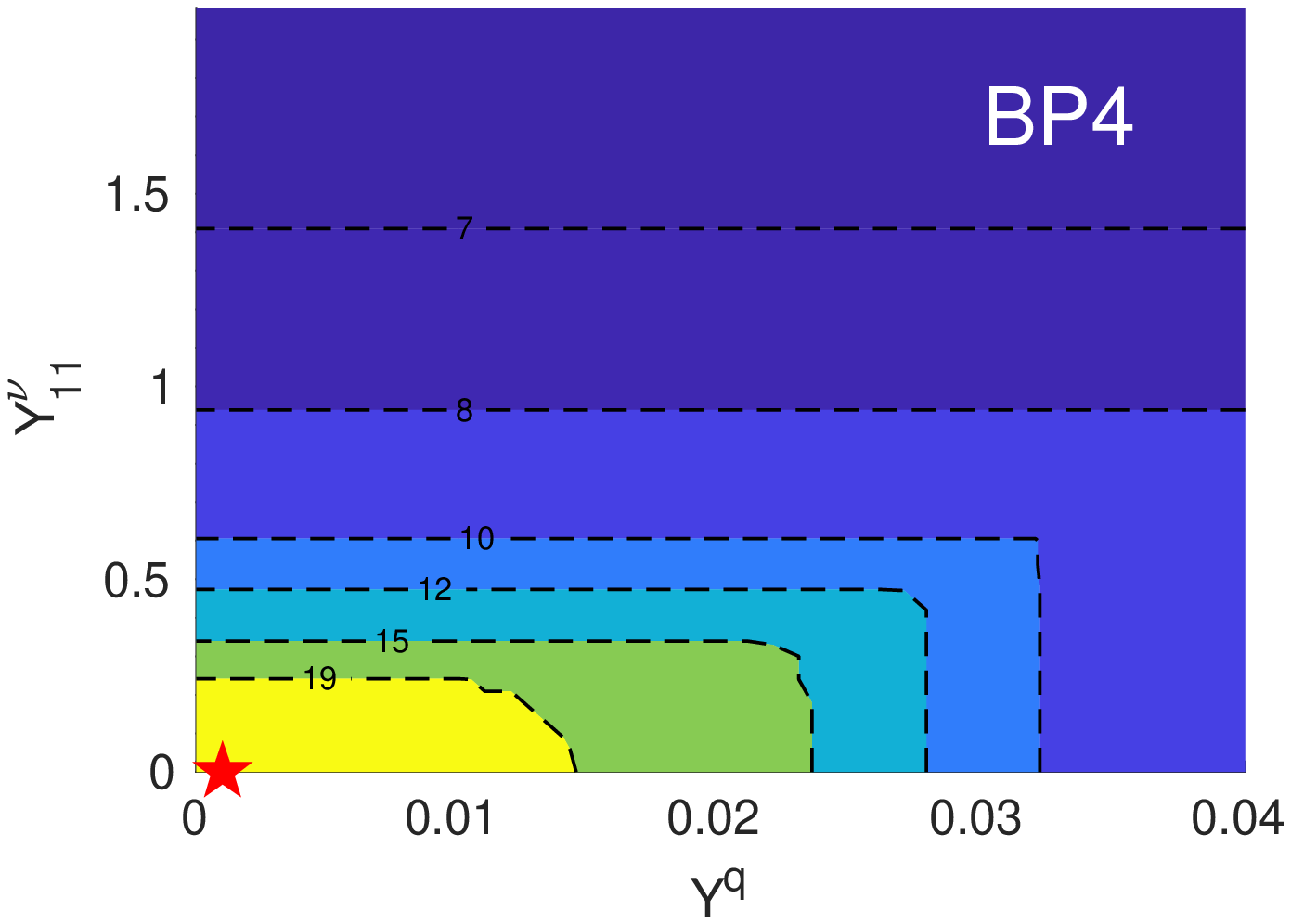}\\
\caption{\label{yf-yq}Vacuum instability scales in $(Y^q,Y^\nu_{11})$ plane in benchmark points \textbf{BP1}-\textbf{BP4}. The red star corresponds to the chosen benchmark point value. The colour coding and the contour numbers are interpreted as in Fig. \ref{fig:bp-mhmt}.}
\label{fig:bp-yy}
\end{figure}

In addition, we have scanned the Dirac neutrino and new quark-like particle Yukawa couplings ($Y^\nu_{11}$ and $Y_Q$, respectively) over $Y^\nu_{11} \in [0,2]$ and $Y_Q \in [0,0.04]$, keeping $Y^\nu_{22}$ and $Y^\nu_{33}$ small, real\footnote{We acknowledge that neutrino Yukawa coupling matrix $Y^\nu$ should be complex in order to allow leptogenesis scenario to work. The vacuum stability analysis, however, is unaffected by this, and we safely ignore the imaginary parts of the Yukawa couplings in this part of the analysis.} and positive but nonzero. See Fig. \ref{fig:bp-yy} for details corresponding to each benchmark point. There we have pointed the area producing a stable vacuum. The Dirac neutrino Yukawa couplings may have a maximum value of $\mathcal{O}(1)$, but a more stringent constraint is found for $Y_Q$. It should be noted that even though from the vacuum instability point of view $Y_Q^{\max} < Y_{11}^{\nu \max}$, this does not imply $Y_Q < Y^\nu_{11}$, since both are in principle free parameters. See Table \ref{bp-results1} for computed values for neutrino masses corresponding to each benchmark. Note that only \textbf{BP4} produces a value of baryon-to-photon ratio comparable to experimental values and a mass of axion consistent with axion dark matter scenario, because it requires axion decay constant $f_A \equiv v_\sigma$ to be $\mathcal{O}(10^{11})$ GeV \cite{Abbott:1982af,Preskill:1982cy,Dine:1982ah}.

\begin{figure}
\centering
\includegraphics[width=0.84\linewidth]{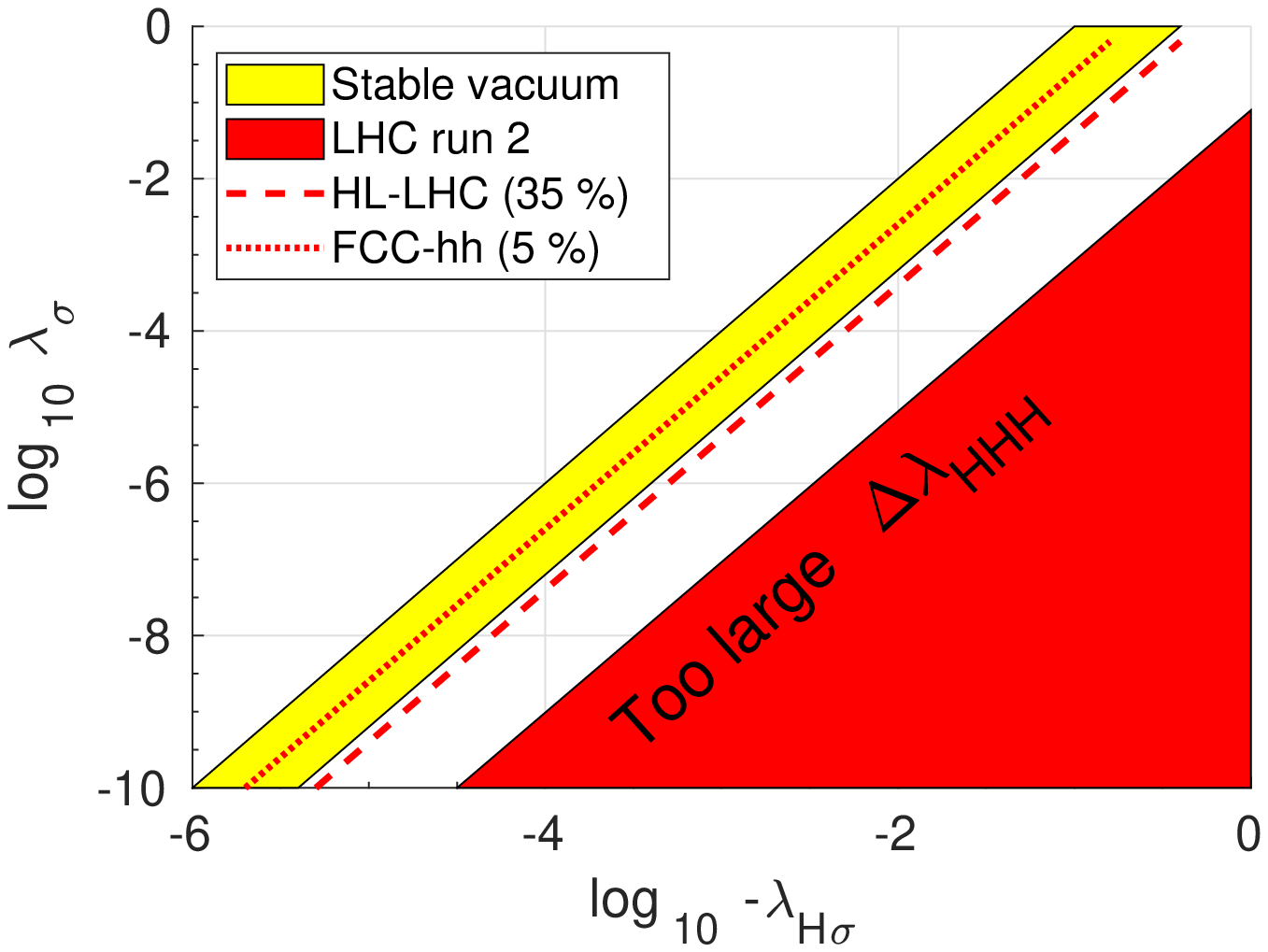}\\
\includegraphics[width=0.84\linewidth]{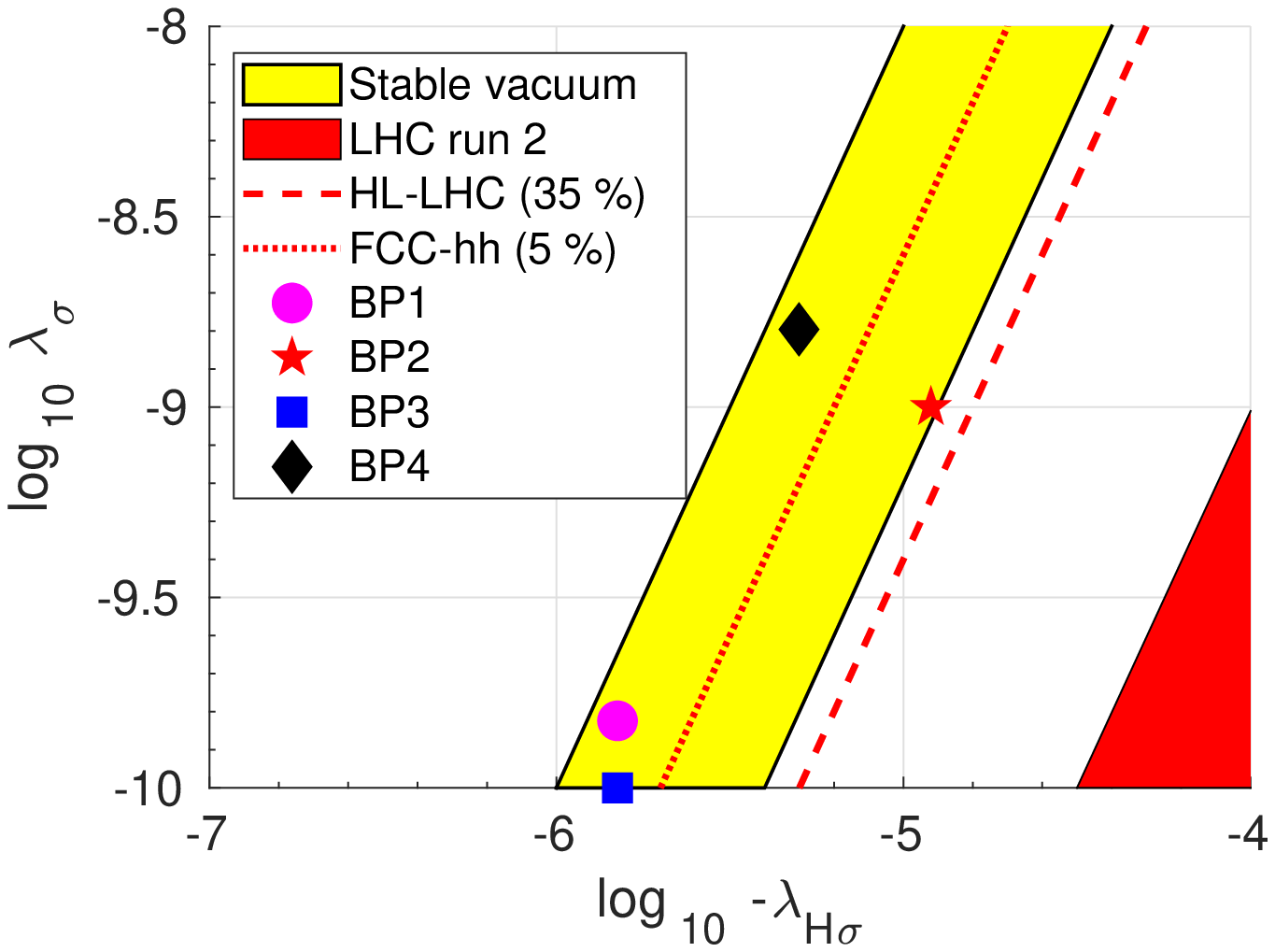}
\caption{\label{fig:triple}\textbf{Above:} Different regions in logarithmic $(-\lambda_{H\sigma},\lambda_{\sigma})$ plane. Yellow band corresponds to stable vacuum configuration. Red area is excluded by second run of the Large Hadron Collider, since the triple Higgs coupling corrections to SMASH would be too large. Dashed line corresponds to the expected sensitivity of high-luminosity LHC and the dotted line to the expected sensitivity of Future Circular Collider on hadronic collision mode. \textbf{Below:} Zoomed-in detail of the figure above, showing in addition our chosen benchmarks.}
\end{figure}

\begin{figure}
\centering
\includegraphics[width=0.836\linewidth]{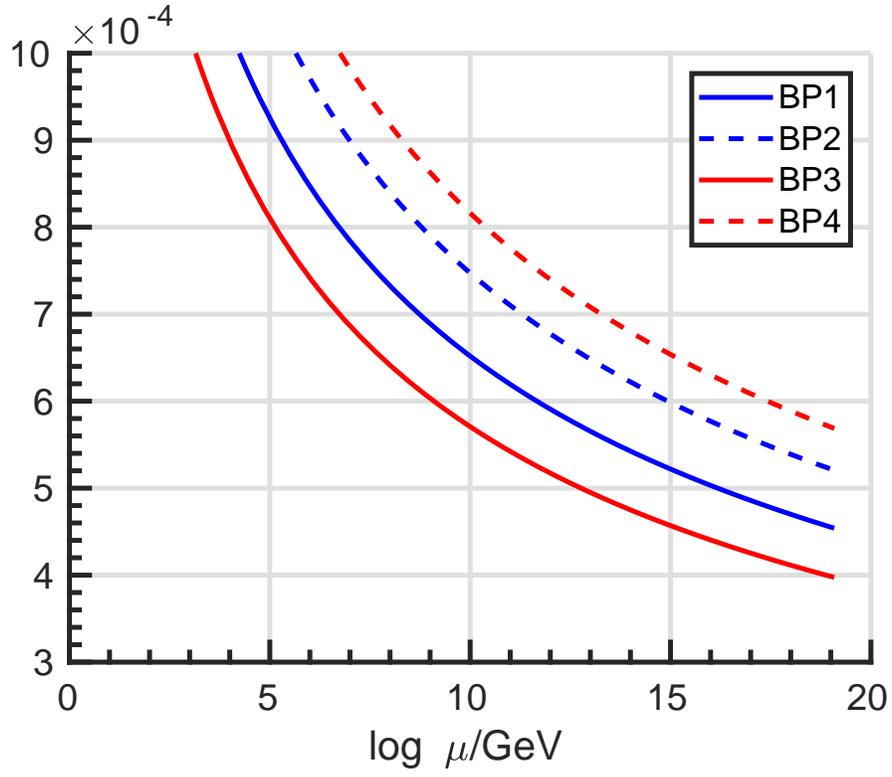}
\caption{\label{fig:yq} Running of $Y_Q$ with different benchmark points.}
\end{figure}

\begin{figure}
\centering
\includegraphics[width=0.836\linewidth]{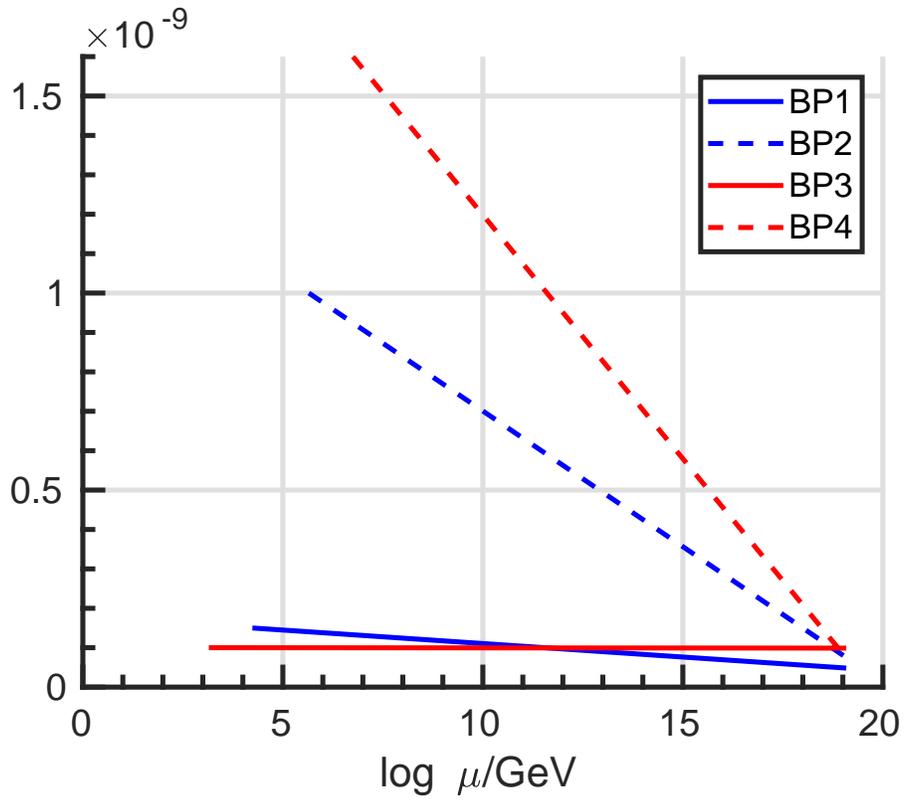}
\caption{\label{fig:lambdas} Running of $\lambda_{\sigma}$ with different benchmark points.}
\end{figure}

\begin{table}[]
\begin{center}
{\renewcommand{\arraystretch}{1.5}\setlength{\tabcolsep}{0.5em}
\begin{tabular}{|c|c|c|c|c|c|}
\hline
\rule{0pt}{3ex}\textbf{Benchmarks} & \textbf{BP1} & \textbf{BP2} & \textbf{BP3} & \textbf{BP4} & \textbf{Experimental values} \\\hline
\rule{0pt}{3ex}$m_1$ (meV) & 23.88 & 0.63 & 0.0055 & 8.90 & \multirow{2}{*}{$\lesssim 55$}\\\cline{1-5}
\rule{0pt}{3ex}$m_2$ (meV) & 25.39 & 8.60 & 8.60 & 12.37& \\ \hline
\rule{0pt}{3ex}$m_3$ (meV) & 55.63 & 50.22 & 50.22 & 51.03& $\lesssim 60$ \\\hline
\rule{0pt}{3ex}$m_1+m_2+m_3$ (meV) & 104.90 & 59.45 & 58.82 & 72.31& $< 120$ \\\hline
\rule{0pt}{3ex}$\Delta m_{21}^2$ ($10^{-5}$ eV$^2$) & 7.41 & 7.36 & 7.39 & 7.39 & 6.79 -- 8.0 \\\hline
\rule{0pt}{3ex}$|\Delta m_{32}^2|$ ($10^{-3}$ eV$^2$)& 2.52 & 2.52 & 2.52 & 2.53 & 2.412 -- 2.625 \\\hline
\rule{0pt}{3ex}$M_1$ (GeV) & $3.27\times 10^6$ & $5.82\times 10^7$ & $1.03\times 10^5$ &$6.73\times 10^8$&\multirow{2}{*}{Unknown} \\\cline{1-5}
\rule{0pt}{3ex}$M_2,M_3$ (GeV) & $9.80\times 10^6$ & $1.74\times 10^8$ & $3.10\times 10^5$ &$2.02\times 10^9$& \\\hline
\end{tabular}}
\caption{\label{bp-results1}The computed values of neutrino masses, sum of light neutrino masses and light neutrino mass squared differences.}
\end{center}
\end{table}

\begin{table}[]
\begin{center}
{\renewcommand{\arraystretch}{1.5}\setlength{\tabcolsep}{0.5em}
\begin{tabular}{|c|c|c|c|c|c|}
\hline
\rule{0pt}{3ex}\textbf{Benchmarks} & \textbf{BP1} & \textbf{BP2} & \textbf{BP3} & \textbf{BP4} & \textbf{Experimental values}\\ \hline
\rule{0pt}{3ex}$\delta$ & 0.017 & 0.144 & 0.023 & 0.016 & None \\\hline 
\rule{0pt}{3ex}$m_A$ (eV) & $5.7\times 10^{-3}$ & $5.7\times 10^{-4}$ & $5.7\times 10^{-2}$ & $5.7\times 10^{-5}$& \multirow{2}{*}{Model-dependent} \\ \cline{1-5}
\rule{0pt}{3ex} $m_\rho$ (GeV) & $1.22\times 10^4$ & $3.16\times 10^5$ & 1000 & $4.00\times 10^6$ & \\\hline 
\rule{0pt}{3ex}$\eta$ & $\sim 10^{-13}$ & $ \sim 10^{-12}$ & $\sim 10^{-15}$ & $\sim 10^{-11}$& $(6.0 \pm 0.2)\times 10^{-10}$\\\hline 
\rule{0pt}{3ex}$\lambda_H(M_{Pl})$ & 0.0070 & 0.4518 & 0.0213 & 0.0048&\multirow{2}{*}{None} \\\cline{1-5}
\rule{0pt}{3ex}$\lambda_S(M_{Pl})$ & $4.83 \times 10^{-11}$& $7.46\times 10^{-11}$& $9.89\times 10^{-11}$& $7.39\times 10^{-11}$ &\\\hline 
\rule{0pt}{3ex} $\Delta\lambda_{HHH}$ & $-2$\% & $-21$\% & $-3$\% & $-2$\% & $<$ 1400\%\\\hline 
\end{tabular}}
\caption{\label{bp-results2}The computed values of threshold correction $\delta$, scalar masses $m_A$ and $m_\rho$, baryon-to-photon ratio $\eta$, quartic self-couplings at $M_{Pl}$, correction to triple Higgs coupling $\Delta \lambda_{HHH}$ compared to the SM prediction.}
\end{center}
\end{table}

%%%%%%%%%%%%%%%%%%%%%%

\subsection{Correction to SM triple Higgs coupling}

%According to PDG, the largest possible experimental value for $\lambda_{HHH}$ is 15 times the SM prediction\footnote{http://pdg.lbl.gov/2019/reviews/rpp2018-rev-higgs-boson.pdf, page 32, chapter 11, section III.4.2.}, from Run 2 data. Only bbar-2gamma ... \textbf{CONTINUE FROM HERE}

The real singlet scalar $\rho$ mixes with the SM Higgs, providing a one-loop correction to SM triple Higgs coupling $\lambda_{HHH}$. We scanned the parameter space with $\log_{10} (-\lambda_{H\sigma}) \in [-7,0]$ and $\log_{10} \lambda_{\sigma} \in [-10,0]$. In each point, we calculated the correction to $\lambda_{HHH}$. See Fig. \ref{fig:triple} for details. We identified section of parameter space excluded by triple Higgs coupling searches from LHC run 2 and determined the area sensitive to future experiments, namely HL-LHC and FCC-hh. We assume HL-LHC uses 14 TeV center-of mass energy and integrated luminosity $\mathcal{L} = 3\, {\rm ab}^{-1}$, for FCC-hh we assume center-of-mass energy 100 TeV and integrated luminosity $\mathcal{L} = 3\, {\rm ab}^{-1}$. The relative correction in Table \ref{bp-results2} is calculated with respect to SM tree-level prediction. We have chosen \textbf{BP2} in a way that its correction to triple Higgs coupling will be observable at FCC-hh \cite{He:2015spf}. Other benchmark points will have such a tiny correction that they will evade the experimental sensitivity of both future experiments.

%%%%%%%%%%%%%%%%%%%%%%

\subsection{Running behaviour of SMASH parameters}

Lastly, we have investigated the running behaviour of the remaining parameters. We found that neutrino Yukawa matrices $Y^\nu$ and $Y^n$, and therefore neutrino masses $m_i$ and $M_i$, mass squared differences $\Delta m_{ij}^2$ and PMNS mixing matrix elements are essentially constant up to the Planck scale. Also, the running of the portal coupling $\lambda_{H\sigma}$ is extremely weak. In contrast, we found the new quark Yukawa coupling to reduce its value in some cases more than 50\%, see Fig. \ref{fig:yq}. Also, the running of the quartic coupling $\lambda_\sigma$ is interesting: in some cases, its value is reduced by more than 90\%. This can be seen from Fig. \ref{fig:lambdas}.

%%%%%%%%%%%%%%%%%%%%%%

\section{\label{conclusions}Conclusions}
 
We have investigated suitable benchmark scenarios on the simplest SMASH model regarding the scalars and neutrinos, constraining the new Yukawa couplings and scalar couplings via the vacuum stability and theory perturbativity requirements. The model can easily account for the neutrino sector, predicting the correct light neutrino mass spectrum while evading the experimental bounds for heavy sterile right-handed Majorana neutrinos. We found an interesting interplay between the triple Higgs coupling correction and SM Higgs quartic coupling correction. Since they are proportional to each other, a large correction to $\lambda_{HHH}$ inevitably leads to large threshold correction. Detecting a $\lambda_{HHH}$ correction larger than $\sim 35\%$ is within the sensitivity of future high-luminosity upgrade of the LHC \cite{Cepeda:2019klc}. If detected, it would, therefore, rule out the simplest scalar sector of the model completely. This would force the model development to nonminimal alternatives, such as an additional scalar doublet or triplet instead of a singlet. These alternatives have been considered by the authors in their recent updated study \cite{Smash3}.

%%%%%%%%%%%%%%%%%%%%%%

\section*{Acknowledgments}

CRD is thankful to Prof. D.I. Kazakov (Director, BLTP, JINR) for support.

%%%%%%%%%%%%%%%%%%%%%%

\bibliography{Smash_031219}{}

%%%%%%%%%%%%%%%%%%%%%%

\end{document}